\documentclass[sn-basic]{sn-jnl}

\usepackage{bm}
\usepackage{fix-cm}
\usepackage{xfrac}
\usepackage{mathrsfs}
\DeclareFontFamily{U}{rsfs}{\skewchar\font127}
\DeclareFontShape{U}{rsfs}{m}{n}{%
   <-6> rsfs5
   <6-8> rsfs7
   <8-> rsfs10
}{}

\usepackage{graphicx}
\usepackage{multirow}
\usepackage{amsmath,amssymb,amsfonts}
\usepackage{amsthm}
\usepackage{mathrsfs}
\usepackage[title]{appendix}
\usepackage{xcolor}
\usepackage{textcomp}
\usepackage{manyfoot}
\usepackage{booktabs}
\usepackage{algorithm}
\usepackage{algorithmicx}
\usepackage{algpseudocode}
\usepackage{listings}

\newcommand{\araa}{Annu. Rev. Astron. Astrophys.} 
\newcommand{\aj}{Astron. J.} 
\newcommand{\apj}{Astrophys. J.} 
\newcommand{\apjs}{Astrophys. J. Suppl. Ser.} 
\newcommand{\ao}{Appl. Opt.} 
\newcommand{\apss}{Astrophys. Space Sci.} 
\newcommand{\aap}{Astron. Astrophys.} 
\newcommand{\cjaa}{Chinese J. Astron. Astrophys.} 
\newcommand{\mnras}{Mon. Not. R. Astron. Soc.} 
\newcommand{\nat}{Nature} 
\newcommand{\pasp}{Publ. Astron. Soc. Pac.} 
\newcommand{\solphys}{Sol. Phys.} 

\graphicspath{{./}{figures/}}

\begin{document}

\title[$\mathrm{H}\alpha$ activity of F-, G-, and K-type stars]{$\mathrm{H}\alpha$ chromospheric activity of F-, G-, and K-type stars observed by the LAMOST Medium-Resolution Spectroscopic Survey}

\author*[1,2]{\fnm{Han} \sur{He}}\email{hehan@nao.cas.cn}

\author[3]{\fnm{Weitao} \sur{Zhang}}

\author[1,2,4]{\fnm{Haotong} \sur{Zhang}}

\author[1,2,4]{\fnm{Song} \sur{Wang}}

\author[1,2,4]{\fnm{Ali} \sur{Luo}}

\author[3]{\fnm{Jun} \sur{Zhang}}

\affil[1]{\orgname{National Astronomical Observatories, Chinese Academy of Sciences}, \orgaddress{\city{Beijing}, \postcode{100101}, \country{China}}}

\affil[2]{\orgname{University of Chinese Academy of Sciences}, \orgaddress{\city{Beijing}, \postcode{100049}, \country{China}}}

\affil[3]{\orgdiv{School of Physics and Optoelectronics Engineering}, \orgname{Anhui University}, \orgaddress{\city{Hefei}, \postcode{230601}, \country{China}}}

\affil[4]{\orgname{CAS Key Laboratory of Optical Astronomy, Chinese Academy of Sciences}, \orgaddress{\city{Beijing}, \postcode{100101}, \country{China}}}

\abstract{
  The distribution of stellar $\mathrm{H}\alpha$ chromospheric activity with respect to stellar atmospheric parameters (effective temperature $T_\mathrm{eff}$, surface gravity $\log\,g$, and metallicity $\mathrm{[Fe/H]}$) and main-sequence/giant categories is investigated for the F-, G-, and K-type stars observed by the LAMOST Medium-Resolution Spectroscopic Survey (MRS).
  A total of 329,294 MRS spectra from LAMOST DR8 are utilized in the analysis.
  The $\mathrm{H}\alpha$ activity index ($I_{\mathrm{H}{\alpha}}$) and the $\mathrm{H}\alpha$ $R$-index ($R_{\mathrm{H}{\alpha}}$) are evaluated for the MRS spectra.
  The $\mathrm{H}\alpha$ chromospheric activity distributions with individual stellar parameters as well as in the $T_\mathrm{eff}$ -- $\log\,g$ and $T_\mathrm{eff}$ -- $\mathrm{[Fe/H]}$ parameter spaces are analyzed based on the $R_{\mathrm{H}{\alpha}}$ index data.
  It is found that:
  (1) for the main-sequence sample, the $R_{\mathrm{H}{\alpha}}$ distribution with $T_\mathrm{eff}$ has a bowl-shaped lower envelope with a minimum at about 6200\,K, a hill-shaped middle envelope with a maximum at about 5600\,K, and an upper envelope continuing to increase from hotter to cooler stars;
  (2) for the giant sample, the middle and upper envelopes of the $R_{\mathrm{H}{\alpha}}$ distribution first increase with a decrease of $T_\mathrm{eff}$ and then drop to a lower activity level at about 4300\,K,
  revealing different activity characteristics at different stages of stellar evolution;
  (3) for both the main-sequence and giant samples, the upper envelope of the $R_{\mathrm{H}{\alpha}}$ distribution with metallicity is higher for stars with $\mathrm{[Fe/H]}$ greater than about $-1.0$,
  and the lowest-metallicity stars hardly exhibit high $\mathrm{H}\alpha$ indices.
  A dataset of $\mathrm{H}\alpha$ activity indices for the LAMOST MRS spectra analyzed is provided with this paper.
}


\maketitle

\section{Introduction} \label{sec:intro}

Stellar chromospheric magnetic activity is believed to widely exist on solar-type stars (e.g., \citealt{2008LRSP....5....2H}).
In the photosphere of stars, the magnetic activity leads to starspots and faculae,
while in the chromosphere immediately above the photosphere, the magnetic activity causes bright plages which can be observed through the line core emission of the chromospheric spectral lines,
such as the H$\alpha$, Ca II H and K, Ca II infrared triplet, etc.
On the Sun and solar-like stars, the chromospheric emission of the H$\alpha$ line is formed at the middle height of the chromosphere \citep{1981ApJS...45..635V};
thus, the H$\alpha$ line can present abundant chromospheric features as demonstrated in the H$\alpha$ monochromatic images of the Sun (e.g., \citealt{1929ApJ....70..265H}),
and has become one of the most commonly used spectral lines for monitoring solar chromospheric activity.
For stars other than the Sun, the H$\alpha$ line is also one of the important approaches to detect and diagnose the activities of stellar chromospheres (e.g., \citealt{2017ARA&A..55..159L}),
and has been included in a variety of spectroscopic observation and sky survey projects.

\citet{1985ApJ...289..269H} evaluated the chromospheric emission flux in the line center of H$\alpha$ line for a sample of F8--G3 main-sequence stars by subtracting a low-activity standard spectrum from the observed spectra.
He found that the chromospheric emission from the H$\alpha$ line correlates linearly with the emission from the Ca II H and K lines,
and the decay of the H$\alpha$ emission flux with stellar age can be fitted by a power law relation.
\citet{1991A&A...251..199P} calibrated the absolute chromospheric flux of H$\alpha$ line for a sample of F8--K5 main-sequence and subgiant stars.
They found that the flux-flux relationship between the H$\alpha$ line and the Ca~II K line of stars is similar to that of solar plages,
and the H$\alpha$ flux of subgiants is typically lower than that of main-sequence stars of the same spectral type.
\citet{2005A&A...431..329L} followed the works of \citet{1985ApJ...289..269H} and \citet{1991A&A...251..199P} and calibrated the chromospheric emission in H$\alpha$ line for a sample of F5--K0 solar neighborhood stars.
They found a well defined age-activity relation until about 2~Gyr by using the stars of open clusters and stellar kinematic groups.

\citet{1990ApJS...74..891R} investigated the H$\alpha$ emission in stellar chromosphere for a sample of main-sequence K- and M-type stars by using the equivalent width of H$\alpha$ line as the chromospheric activity indicator.
Their results confirm the presence of lower and upper limits to the H$\alpha$ equivalent width for a given spectral type of stars,
and both the lower and upper limits increase toward cooler stars.
\citet{2017ApJ...834...85N} used the H$\alpha$ equivalent width to investigate the chromospheric activity of M-type main-sequence stars and its relation to the stellar rotation period.
They found that the active and inactive M-type stars can be separated by a threshold in the mass-period parameter space, which indicates that the activity of H$\alpha$ line can be a useful approach for diagnosing rotation periods of stars.
\citet{2020MNRAS.495.1252Z, 2021ApJS..253...19Z} employed the equivalent width of H$\alpha$ line to analyze the variations of stellar chromospheric activity based on the H$\alpha$ spectral data from LAMOST Data Release 7.

The H$\alpha$ chromospheric activity of stars can also be indicated by the activity index of H$\alpha$ line,
like the $S$-index introduced for the Ca II H and K lines \citep{1968ApJ...153..221W, 1978PASP...90..267V, 1991ApJS...76..383D, 1995ApJ...438..269B}, which is defined as the ratio of the mean flux of the H$\alpha$ line core emission to the mean flux of two continuum bands on the two sides of the line.
\citet{2003A&A...403.1077K} adopted the H$\alpha$ activity index as the indicator of the chromospheric activity of Barnard's star, 
a M4 main-sequence star that is just 1.82 pc away from the solar system.
\citet{2007A&A...474..293B} used the H$\alpha$ activity index as one of the activity parameters to indicate the chromospheric activity of another nearby main-sequence star GJ 674 (spectral type M2.5; distance about 4.5 pc) for characterizing the exoplanet of the star.
\citet{2009A&A...495..959B} employed the H$\alpha$ activity index as one of the activity indicators to investigate the activity properties of a K-type exoplanet host star, HD 189733.
\citet{2011A&A...534A..30G} adopted the H$\alpha$ activity index as one of the activity measures to investigate the long-term magnetic activities of a sample of M-type main-sequence stars from the HARPS M-dwarf planet search program.
\citet{2013ApJ...764....3R} utilized the H$\alpha$ activity index to investigate the stellar cycles and mean activity levels for a sample of K5--M5 stars in solar neighborhood.
They also analyzed the influence of the stellar activities revealed by the H$\alpha$ variation to the precision of the radial velocity measurements for the exoplanet host stars.
\citet{2016ApJ...832..112R} used the H$\alpha$ activity index as one of the tracers of stellar magnetic activity to examine the activity sensitivity of eight near-infrared atomic lines for detecting exoplanets around mid- to late M stars through the radial velocity approach.

In order to compare H$\alpha$ activity between stars of different spectral types, 
a better indicator of stellar H$\alpha$ chromospheric activity than the H$\alpha$ equivalent width or the H$\alpha$ activity index is the ratio of the stellar H$\alpha$ luminosity to the stellar bolometric luminosity (e.g., \citealt{2004PASP..116.1105W, 2011AJ....141...97W}),
like the $R$-index employed for the Ca II H and K lines (e.g., \citealt{1982A&A...107...31M, 1984ApJ...276..254H, 1984ApJ...279..763N}).
\citet{2004PASP..116.1105W} introduced a $\chi$ factor to help evaluate the $R$-index of H$\alpha$ line, which is defined as the ratio of the stellar continuum flux near H$\alpha$ line to the stellar bolometric flux; 
the $R$-index of H$\alpha$ line can then be obtained by multiplying the $\chi$ factor by the measured equivalent width of H$\alpha$ line.

The LAMOST (Large Sky Area Multi-Object Fiber Spectroscopic Telescope, also named Guoshoujing Telescope; \citealt{2012RAA....12.1197C, 2012RAA....12..723Z}) Medium-Resolution Spectroscopic Survey (MRS) has been observing the spectral data of H$\alpha$ line of stars since 2017.
The spectral resolution power ($R=\lambda/\delta\lambda$) of the MRS data is about 7500 \citep{2020arXiv200507210L}.
The full wavelength range of the H$\alpha$ line as well as the adjacent continuum is well covered by MRS spectra.
After several years of observation, the LAMOST MRS has acquired millions of spectra of celestial objects;
and more than a million spectra are provided with stellar atmospheric parameters, such as effective temperature ($T_\mathrm{eff}$), surface gravity ($\log\,g$), and metallicity ($\mathrm{[Fe/H]}$).
The large amount of MRS spectra is well suited for studying the overall distribution of stellar H$\alpha$ chromospheric activity \citep{2021RNAAS...5....6H}.

In this paper, we investigate the distribution of the stellar H$\alpha$ chromospheric activity with respect to stellar atmospheric parameters and main-sequence/giant categories by using the H$\alpha$ line spectral data of MRS from LAMOST Data Release 8 (DR8).
We employ the H$\alpha$ activity index and the H$\alpha$ $R$-index as the indicators of H$\alpha$ chromospheric activity.
The spectral sample analyzed includes the MRS spectra of F-, G-, and K-type stars, in which the spectra of main-sequence stars and giants are distinguished.
The H$\alpha$ chromospheric activity distributions with individual stellar parameters as well as in the $T_\mathrm{eff}$ -- $\log\,g$ and $T_\mathrm{eff}$ -- $\mathrm{[Fe/H]}$ parameter spaces are analyzed and compared for the main-sequence and giant spectral samples .

It should be mentioned that the LAMOST Low-Resolution Spectroscopic Survey (LRS) also involves the H$\alpha$ line and can be utilized for studying the stellar H$\alpha$ chromospheric activity property (e.g., \citealt{2016A&A...594A..39F, 2021ApJS..253...19Z}).
The spectral resolution power of the LRS data is about 1800 (\citealt{2012RAA....12..723Z}), which is lower than that of the MRS data.
Considering that the chromospheric emission mainly comes from the narrow band at the center of the H$\alpha$ line, the MRS spectra have an advantage over the LRS spectra in resolving the line core emission.
In this paper, we focus on the MRS data to investigate stellar H$\alpha$ chromospheric activity.

The content of the paper is structured as follows.
In Section \ref{sec:MRS_observations}, we explain the H$\alpha$ line observations by the LAMOST MRS.
In Section \ref{sec:spectral_sample_selection}, we describe the selection procedure for the MRS spectral sample used in the analysis.
In section \ref{sec:halpha_indices}, we give the specific definitions of the H$\alpha$ activity index and the H$\alpha$ $R$-index adopted in this work, and evaluate the values of the H$\alpha$ indices for the selected MRS spectra.
In Section \ref{sec:results}, we analyze in detail the distribution of H$\alpha$ chromospheric activity with stellar atmospheric parameters and main-sequence/giant categories based on the H$\alpha$ $R$-index data.
In Section \ref{sec:dataset}, we describe the dataset of the H$\alpha$ activity indices obtained in this work.
Section \ref{sec:conclusion} is the conclusion and discussion.

\section{$\mathrm{H}\alpha$ line observations by the LAMOST MRS} \label{sec:MRS_observations}

The MRS of LAMOST collects spectra of celestial objects in a blue band (4950--5350\,{\AA}) and a red band (6300--6800\,{\AA}) simultaneously \citep{2020arXiv200507210L}.
The H$\alpha$ line is included in the red band.
The duration of a single exposure of MRS is about 20 minutes.
An observed object is usually exposed multiple times in succession by the MRS.
The coadded spectrum of the object can then be generated by combining the multiple single-exposure spectra.
All the single-exposure spectra and the coadded spectrum of the observed object in one observation are stored in a same FITS data file and share a same MRS {\tt\string obsid} (MRS unique observation identifier).
In this work, we use the coadded spectral data of LAMOST MRS to analyze stellar H$\alpha$ chromospheric activity,
for a coadded spectrum generally has a higher signal-to-noise ratio than the associated individual single-exposure spectrum.

\begin{figure}
  \centering
  \includegraphics[width=1.02\textwidth]{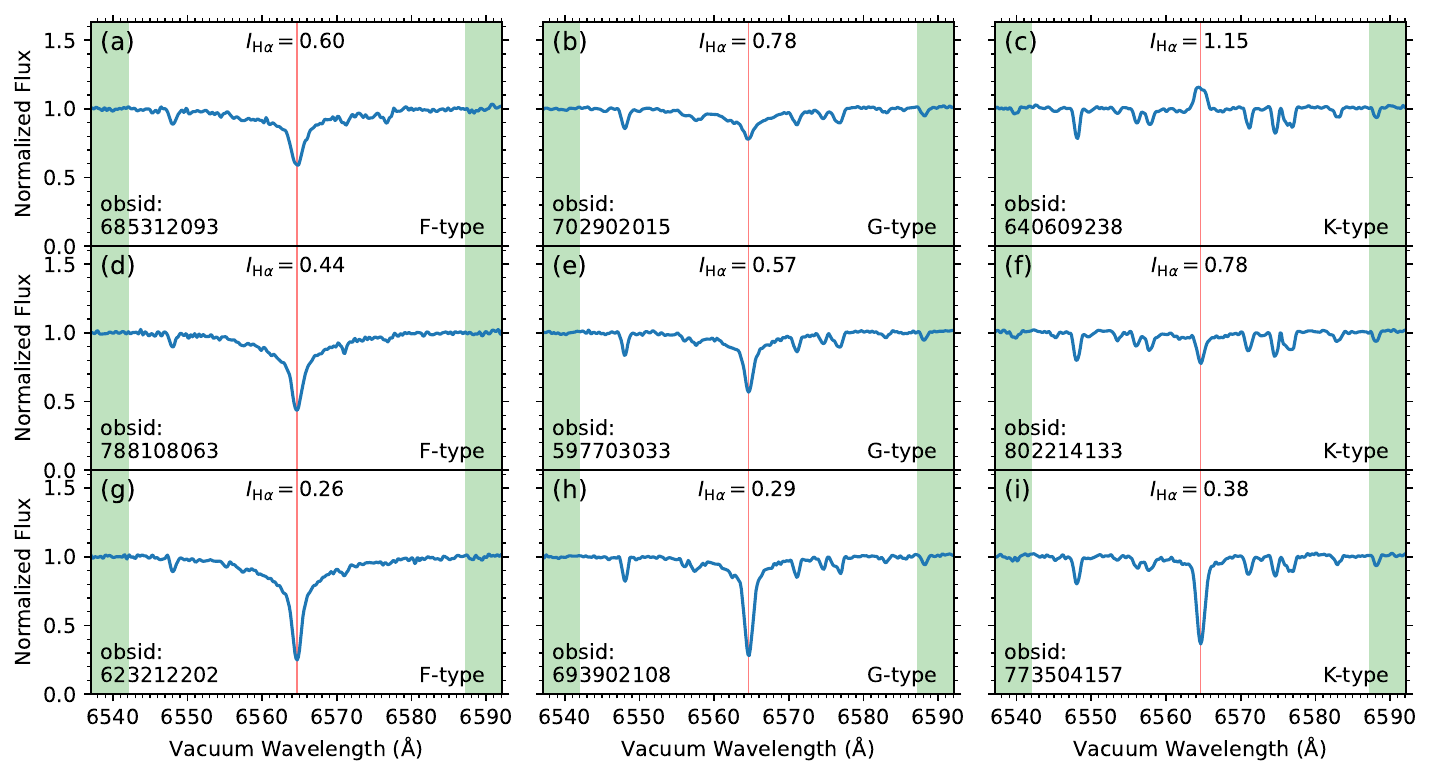}
  \caption{Example of the H$\alpha$ line observations by the LAMOST MRS.
    The H$\alpha$ spectra in the three columns from left to right belong to the F-, G-, and K-type main-sequence stars, respectively.
    In each column, the three H$\alpha$ spectra from bottom to top represent the increasing H$\alpha$ activity levels.
    The wavelengths of the spectra have been transformed to the values in the rest frame.
    The fluxes of the spectra have been normalized by the continuum values.
    The center band (shown in red) and the two continuum bands (shown in green) adopted in this work for the H$\alpha$ activity index ($I_{\mathrm{H}{\alpha}}$) evaluation (see Section \ref{sec:I_index}) are indicated.
    The MRS {\tt\string obsid} (MRS unique observation identifier) and the evaluated $I_{\mathrm{H}{\alpha}}$ value are given for each of the example spectra.}
  \label{fig:example_mrs_halpha_spectra}
\end{figure}

Figure \ref{fig:example_mrs_halpha_spectra} gives an example of the H$\alpha$ line observations by the LAMOST MRS.
The H$\alpha$ spectra in the three columns of Figure \ref{fig:example_mrs_halpha_spectra} (from left to right) belong to the F-, G-, and K-type  main-sequence stars, respectively.
In each column, the three H$\alpha$ spectra from bottom to top represent the increasing H$\alpha$ activity levels.
As shown in Figure \ref{fig:example_mrs_halpha_spectra},
LAMOST uses vacuum wavelength in the released spectral data.
The vacuum wavelength of H$\alpha$ line center in the rest frame is 6564.614\,{\AA} as adopted by the Sloan Digital Sky Survey (SDSS; \citealt{2002AJ....123..485S}),
which can be converted to and from the air wavelength of H$\alpha$ line (6562.801\,{\AA}) using the formula given by \citet{1996ApOpt..35.1566C}.

\section{Selection of MRS spectral sample} \label{sec:spectral_sample_selection}

We utilize the MRS spectra of F-, G-, and K-type stars from LAMOST DR8 v1.1\footnote{\url{http://www.lamost.org/dr8/v1.1/}} for the H$\alpha$ chromospheric activity analysis.
The MRS data in LAMOST DR8 were observed from September 2017 to June 2020 (three full observation years).
The {\tt LAMOST MRS Parameter Catalog} of DR8 contains 1,230,307 coadded spectra of stars with determined stellar atmospheric parameters ($T_\mathrm{eff}$, $\log\,g$, and $\mathrm{[Fe/H]}$).
The MRS spectral sample analyzed in this work are selected from this catalog.

The spectra of F-, G-, and K-type stars are selected from the {\tt LAMOST MRS Parameter Catalog} by using the effective temperature condition of $4000\,{\rm K} < T_\mathrm{eff} < 7000\,{\rm K}$ \citep{1966ARA&A...4..193J}.
We use the effective temperature parameter {\tt\string teff\_lasp} included in the catalog (and the parameters {\tt\string logg\_lasp} and {\tt\string feh\_lasp} for the surface gravity and metallicity, respectively, in the following analysis) to perform the spectral sample selection,
which is provided by the LAMOST Stellar Parameter Pipeline (LASP; \citealt{2015RAA....15.1095L}).
The number of the MRS spectra in the catalog that meet the above $T_\mathrm{eff}$ condition is 1,148,919.

\begin{figure}
  \centering
  \includegraphics[width=0.58\textwidth]{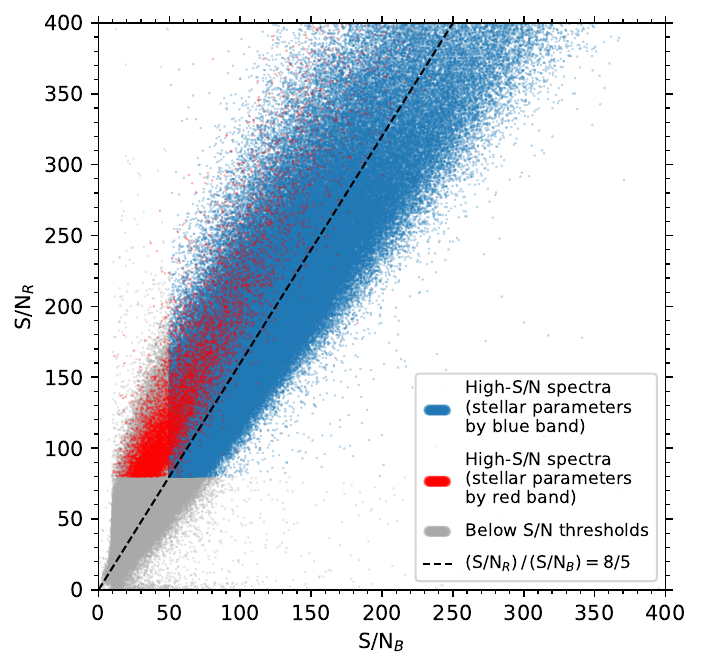}
  \caption{Scatter diagram of ${\rm S/N}_B$ versus ${\rm S/N}_R$ for the MRS spectra of F-, G-, and K-type stars ($4000\,{\rm K} < T_\mathrm{eff} < 7000\,{\rm K}$) in the {\tt LAMOST MRS Parameter Catalog}.
  The dashed line indicates the ratio of $({\rm S/N}_R) / ({\rm S/N}_B) = 8/5$.
  The high-S/N spectra that meet the S/N conditions (see main text for details) are highlighted in blue and red for the samples whose stellar atmospheric parameters are determined by the blue band and red band of MRS, respectively.
  The spectral sample below the S/N thresholds is displayed in gray.
  Note that the maximum S/N value of the MRS spectra is 1000;
  the range of the axes is limited to 400 to better illustrate the S/N thresholds for selection of the high-S/N spectra.
  }
  \label{fig:scatter_diagram_snb_vs_snr}
\end{figure}

LAMOST provides separate signal-to-noise ratio (S/N) parameters for the blue and red bands of MRS (denoted by ${\rm S/N}_B$ and ${\rm S/N}_R$, respectively),
which are the median of the S/N values of all data points in a band.
We select the spectra with high S/N values for our analysis,
as a higher S/N generally corresponds to a smaller uncertainty of spectral fluxes and more accurate stellar atmospheric parameters determined from the spectra \citep{2015RAA....15.1095L}.
In Figure \ref{fig:scatter_diagram_snb_vs_snr}, we show a scatter diagram of ${\rm S/N}_B$ versus ${\rm S/N}_R$\footnote{The values of ${\rm S/N}_B$ and ${\rm S/N}_R$ are from the {\tt LAMOST MRS General Catalog}.} for the spectra of F-, G-, and K-type stars in the {\tt LAMOST MRS Parameter Catalog} to illustrate the selection of the high-S/N spectra.

Since the H$\alpha$ line is in the red band of MRS, to select the high-S/N spectra, we first introduce an S/N condition of ${\rm S/N}_R \geq 80.0$ for the red band data of MRS.
The threshold of ${\rm S/N}_R$ is determined empirically to ensure an appropriate upper limit of the uncertainties of the evaluated H$\alpha$ activity index values (see Section \ref{sec:I_index}).
On the other hand, there are only a small fraction of the MRS spectra whose stellar atmospheric parameters are determined by the red band of MRS,
and the stellar parameters of most MRS spectra are determined based on the blue band of MRS (see the data release documents of LAMOST for more details).
For the spectra whose stellar parameters are determined by the blue band, in addition to the red band S/N condition,
we introduce another S/N condition for the blue band data of MRS, which is ${\rm S/N}_B \geq 50.0$.
The threshold of ${\rm S/N}_B$ is determined from the ${\rm S/N}_R$ threshold by considering that, for the MRS spectra of F-, G-, and K-type stars, the ratio of $({\rm S/N}_R)/({\rm S/N}_B)$ is roughly 8/5 as demonstrated in Figure \ref{fig:scatter_diagram_snb_vs_snr}.

From the {\tt LAMOST MRS Parameter Catalog}, we find 417,935 MRS spectra of F-, G-, and K-type stars that meet the above S/N conditions,
in which the stellar parameters of 408,309 spectra are determined by the blue band of MRS, and the stellar parameters of 9,626 spectra are determined by the red band of MRS.
The two sets of samples are highlighted in blue and red, respectively, in Figure \ref{fig:scatter_diagram_snb_vs_snr}.

{\tt LAMOST MRS Parameter Catalog} also provides the projected rotational velocity ($v{\sin}i$) values determined by the LASP (parameter {\tt\string vsini\_lasp} in the catalog) for the spectra with $v{\sin}i > 30$\,\rm km/s.
The rotational broadening of H$\alpha$ line caused by a large $v{\sin}i$ can affect the evaluated values of the H$\alpha$ activity index (see Section \ref{sec:I_index} and Appendix \ref{sec:diff_I_index_vs_vsini}).
To minimize the effects of rotational broadening,
we only keep the spectra with $v{\sin}i$ less than 30\,km/s, i.e., the spectra whose {\tt\string vsini\_lasp} values are not available in the catalog (indicated by a value of -9999.0).
The number of the selected spectra after this condition is 337,742.
Those spectra with $v{\sin}i$ greater than 30\,km/s (80,193 in total) are removed from our sample.

Some of the MRS spectra have invalid flux values in the wavelength range of H$\alpha$ line (e.g., flux equaling to or less than zero) or missing auxiliary parameters in the LAMOST MRS catalog (e.g., the parameter of radial velocity used in Section \ref{sec:I_index});
a small portion of the MRS spectra comes from the stellar objects close to or in the Galactic nebulae, 
which belong to the LAMOST medium-resolution spectral survey of Galactic Nebulae (MRS-N), a sub-project of LAMOST MRS (see \citealt{2021RAA....21...96W, 2022RAA....22g5015W} for more details).
These spectra are also removed from our sample.
We finally get 329,294 MRS spectra of F-, G-, and K-type stars for the following H$\alpha$ chromospheric activity analysis.

\begin{figure}
  \centering
  \includegraphics[width=0.56\textwidth]{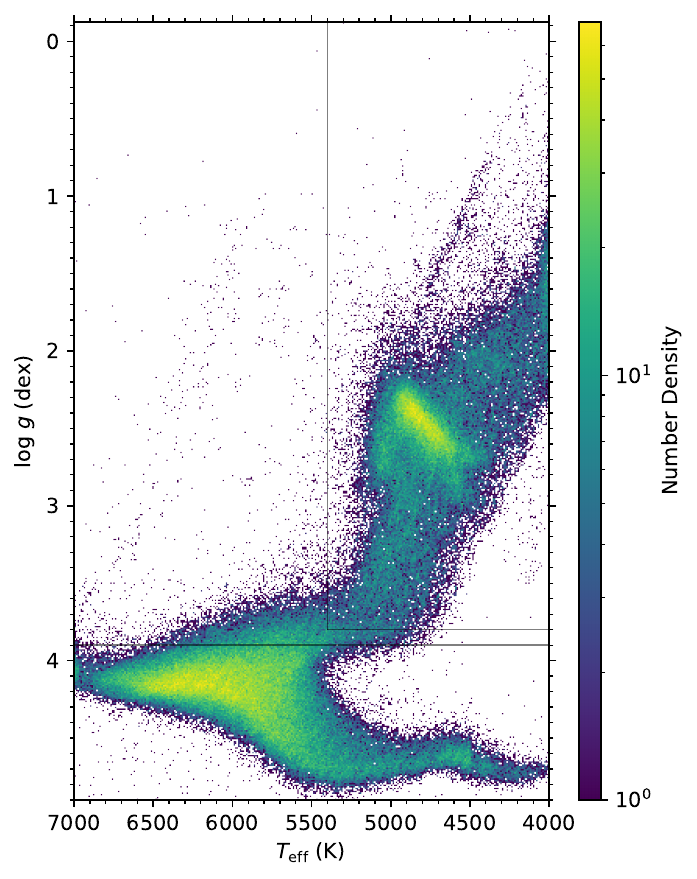}
  \caption{Distribution of the selected MRS spectra of F-, G-, and K-type stars in the $T_\mathrm{eff}$ -- $\log\,g$ parameter space.
  Color scale indicates number density of the spectral sample.
  The horizontal and vertical lines (corresponding to $\log\,g = 3.9$, $\log\,g = 3.8$, and $T_\mathrm{eff} = 5400\,{\rm K}$, respectively, from bottom to top) are the dividing lines for the main-sequence sample (bottom region) and the giant sample (upper-right region).}
  \label{fig:number_density_teff_vs_logg}
\end{figure}

Because the main-sequence and giant stars are at different evolution stages and might present different chromospheric activity properties (e.g., \citealt{1991A&A...251..199P, 2005A&A...431..329L}), we distinguish the two stellar categories in the analysis.
To illustrate the partitioning of the main-sequence and giant samples,
in Figure \ref{fig:number_density_teff_vs_logg} we show the distribution of the selected MRS spectra of F-, G-, and K-type stars in the $T_\mathrm{eff}$ -- $\log\,g$ parameter space, with number density of the spectral sample being indicated by color scale.
The horizontal and vertical lines in Figure \ref{fig:number_density_teff_vs_logg} are the dividing lines for the main-sequence and giant samples,
whose position is determined by referring to the stellar parameter ranges of main-sequence and giant stars in the literature (e.g., \citealt{1981Ap&SS..80..353S, 2019A&A...624A..19B, 2019LRSP...16....4G, 2019FrASS...6...52K, 2020MNRAS.495.1252Z, 2022ApJ...929..124C}).
As shown in Figure \ref{fig:number_density_teff_vs_logg},
we use the condition of $\log\,g \ge 3.9$ for the main-sequence sample (bottom region in Figure \ref{fig:number_density_teff_vs_logg}),
and the condition of $T_\mathrm{eff} < 5400\,{\rm K}$ and $\log\,g < 3.8$ for the giant sample (upper-right region in Figure \ref{fig:number_density_teff_vs_logg}).
The main-sequence and giant regions are separated by an intermediate zone to ensure the purity of the two samples.
In the selected MRS spectra of F-, G-, and K-type stars,
the numbers of the main-sequence spectra and the giant spectra that meet the above conditions are 188,617 and 121,740, respectively.
Although not included in the main-sequence or giant sample,
the spectra in the intermediate zone (18,937 in total) are still used in the analysis.

Because the stellar atmospheric parameters determined by the LASP ({\tt\string teff\_lasp}, {\tt\string logg\_lasp}, and {\tt\string feh\_lasp} in the {\tt LAMOST MRS Parameter Catalog}) are extensively used in our analysis,
to rule out the possibility of degeneracy between the stellar parameter values,
we examine the distribution of the $\mathrm{[Fe/H]}$ values in the $T_\mathrm{eff}$ -- $\log\,g$ parameter space for the selected MRS spectra of  F-, G-, and K-type stars (see Appendix \ref{sec:feh_in_teff_logg_space} for details).
The result does not suggest a degeneracy (apparent correlation) between the values of $\mathrm{[Fe/H]}$ and the values of $T_\mathrm{eff}$ (or $\log\,g$),
which demonstrates the reliability of the stellar atmospheric parameter values.

\section{$\mathrm{H}\alpha$ indices of stellar chromospheric activity} \label{sec:halpha_indices}

\subsection{$\mathrm{H}\alpha$ activity index} \label{sec:I_index}

We employ the H$\alpha$ activity index (denoted by $I_{\mathrm{H}{\alpha}}$) as the primary index of stellar H$\alpha$ chromospheric activity,
for it directly relies on the observed spectral flux data of MRS.
The H$\alpha$ activity index is defined as the ratio of the mean flux of the center band of H$\alpha$ line to the mean flux of two continuum bands on the two side of H$\alpha$ line, 
i.e. (e.g., \citealt{2003A&A...403.1077K, 2013ApJ...764....3R}),
\begin{equation} \label{equ:I_halpha_index_definition}
  I_{\mathrm{H}{\alpha}} = \frac{\overline{F}_{\mathrm{H}{\alpha}}}{\overline{F}_\mathrm{cont}} = \frac{\overline{F}_{\mathrm{H}{\alpha}}}{(\overline{F}_\mathrm{cont, 1}+\overline{F}_\mathrm{cont, 2})/2},
\end{equation}
where $F$ represents the observed flux data of MRS, $\overline{F}_{\mathrm{H}{\alpha}}$ is the mean flux of the H$\alpha$ center band,
$\overline{F}_\mathrm{cont, 1}$ and $\overline{F}_\mathrm{cont, 2}$ are the mean fluxes of the continuum bands on the blue side and red side of H$\alpha$ line, respectively,
and $\overline{F}_\mathrm{cont}=(\overline{F}_\mathrm{cont, 1}+\overline{F}_\mathrm{cont, 2})/2$ is the mean flux of the both continuum bands.
The value of $I_{\mathrm{H}{\alpha}}$ defined by Equation (\ref{equ:I_halpha_index_definition}) is in the range of 0.0--1.0 for the H$\alpha$ absorption line profile and greater than 1.0 for the H$\alpha$ emission line profile (see example in Figure \ref{fig:example_mrs_halpha_spectra}).

To evaluate the H$\alpha$ activity index based on Equation (\ref{equ:I_halpha_index_definition}), \citet{2003A&A...403.1077K}, \citet{2007A&A...474..293B}, and \citet{2009A&A...495..959B} adopted a center bandwidth of H$\alpha$ line of 0.678\,{\AA},
while \citet{2011A&A...534A..30G} and \citet{2013ApJ...764....3R, 2016ApJ...832..112R} used a 1.6\,{\AA} wide center band.
The two continuum bands chosen by these works are centered at 6550.87\,{\AA} and 6580.31\,{\AA} (wavelengths in air) with bandwidths of 10.75\,{\AA} and 8.75\,{\AA}, respectively (\citealt{2003A&A...403.1077K, 2011A&A...534A..30G}),
which are optimized and hence more suitable for the relatively narrow H$\alpha$ line profiles of K- and M-type stars.

In this work, we extend the H$\alpha$ activity index analysis to F- and G-type stars.
To match the relatively wider H$\alpha$ line profiles of these types of stars
as demonstrated in Figure \ref{fig:example_mrs_halpha_spectra},
we adopt a different definition of the center and continuum bands of H$\alpha$ line than that for the K- and M-type stars described above.
In our definition, the width of the center band of H$\alpha$ line is set to 0.25\,{\AA},
which is commonly used by the solar H$\alpha$ filtergram observations (e.g., \citealt{2007ChJAA...7..281Z, 2013RAA....13.1509F, 2014RAA....14..705L, 2017SoPh..292...63I}) to minimize the contamination of photospheric radiation from the line wings.
The center wavelengths of the two continuum bands are both 25\,{\AA} away from the line center of H$\alpha$ line,
and the width of each continuum band is set to 5\,{\AA}.
A diagram illustration of the center and continuum bands adopted in this work for the $I_{\mathrm{H}{\alpha}}$ evaluation has been given in Figure \ref{fig:example_mrs_halpha_spectra}.

Owing to the radial velocities of the observed stellar objects,
the wavelength values in the original MRS spectral data need to be transformed to the wavelengths in the rest frame before $I_{\mathrm{H}{\alpha}}$ evaluation.
We adopt the radial velocity parameter {\tt rv\_r0} included in the {\tt LAMOST MRS Parameter Catalog} to perform the wavelength transformation,
which is determined based on the red band spectra of MRS by cross-correlation with the template spectra (see the data release documents of LAMOST for more details).

\begin{figure}
  \centering
  \includegraphics[width=1.02\textwidth]{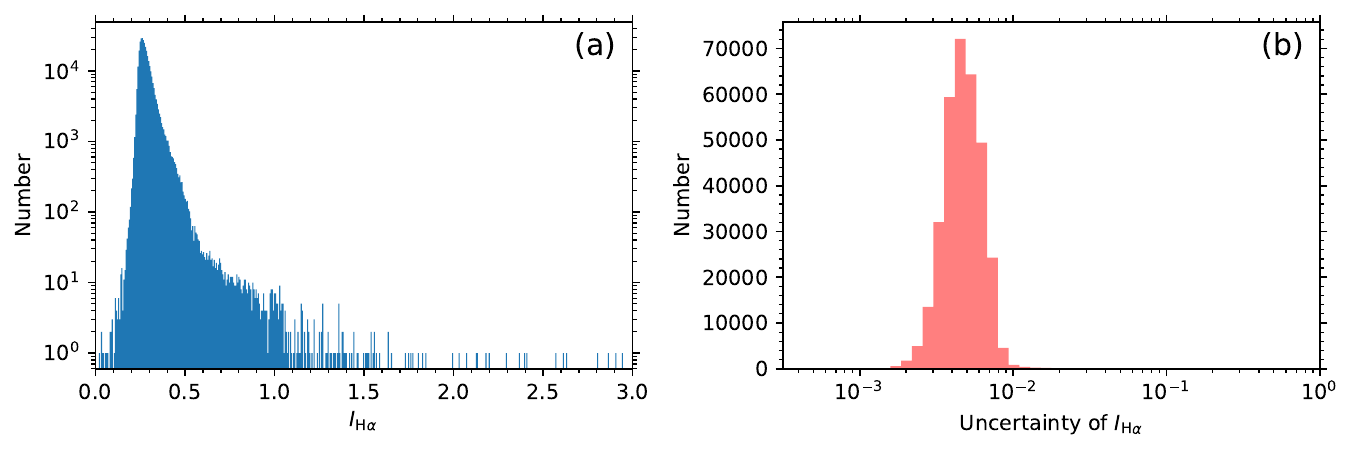}
  \caption{Histograms of the evaluated $I_{\mathrm{H}{\alpha}}$ index values (panel a) and their uncertainties (panel b) for the selected MRS spectra of F-, G-, and K-type stars.}
  \label{fig:histogram_I_halpha_and_err}
\end{figure}

We calculate the values of $I_{\mathrm{H}{\alpha}}$ using Equation (\ref{equ:I_halpha_index_definition}) for all the MRS spectra of F-, G-, and K-type stars selected in Section \ref{sec:spectral_sample_selection}.
The stochastic uncertainties of the $I_{\mathrm{H}{\alpha}}$ values (denoted by $\delta I_{\mathrm{H}{\alpha}}$) are also estimated,
which takes into account the spectral flux uncertainty, the radial velocity uncertainty, and the discretization in the spectral data (\citealt{2022ApJS..263...12Z}).
Figure \ref{fig:histogram_I_halpha_and_err} shows the histograms of the evaluated $I_{\mathrm{H}{\alpha}}$ index values and the estimated uncertainties of $I_{\mathrm{H}{\alpha}}$.
It can be seen from Figure \ref{fig:histogram_I_halpha_and_err} that most $I_{\mathrm{H}{\alpha}}$ values are distributed in the range of less than about 1.7,
and the uncertainties of $I_{\mathrm{H}{\alpha}}$ are on the order of magnitude of $10^{-3}$ to $10^{-2}$.
The data of the obtained $I_{\mathrm{H}{\alpha}}$ index values and their  uncertainties are available in the online dataset of this paper (see Section \ref{sec:dataset}).

It needs to be noted that the projected rotational velocity, $v{\sin}i$, can distort the shape of the H$\alpha$ line owing to the rotational broadening effect (e.g., \citealt{1985ApJ...289..269H, 1991A&A...251..199P}),
and hence leads to an increment of $I_{\mathrm{H}{\alpha}}$ values (denoted by $\Delta I_{\mathrm{H}{\alpha}}$) which takes a positive value for the absorption line profile and negative value for the emission line profile.
For this reason, we have removed the spectra with $v{\sin}i$ greater than 30\,km/s from our sample, as described in Section \ref{sec:spectral_sample_selection}.
LAMOST MRS catalog does not provide $v{\sin}i$ values for the spectra with $v{\sin}i$ less than 30\,km/s,
therefore, 30\,km/s can be regarded as the upper limit of $v{\sin}i$ for the MRS spectral sample analyzed in this work.

In Appendix \ref{sec:diff_I_index_vs_vsini}, we estimate the relationship between the magnitudes of $v{\sin}i$ and $\Delta I_{\mathrm{H}{\alpha}}$ by using the nine example spectra shown in Figure \ref{fig:example_mrs_halpha_spectra}.
The results are plotted in Figure \ref{fig:I_index_increment_vs_vsini}.
It can be seen from Figure \ref{fig:I_index_increment_vs_vsini} that for the $v{\sin}i$ values less than 30\,km/s, the absolute magnitude of $\Delta I_{\mathrm{H}{\alpha}}$ is generally below about $0.02 \thicksim 0.05$ for higher $I_{\mathrm{H}{\alpha}}$ values (panels a--f of Figure \ref{fig:I_index_increment_vs_vsini}) and below about $0.06 \thicksim 0.08$ for lower $I_{\mathrm{H}{\alpha}}$ values (panels g--i of Figure \ref{fig:I_index_increment_vs_vsini}).
The spectra with a lower $I_{\mathrm{H}{\alpha}}$ tend to have a larger $\Delta I_{\mathrm{H}{\alpha}}$ owing to their deeper spectral line profile (see panels g--i of Figure \ref{fig:example_mrs_halpha_spectra}).
On the other hand, stars with a lower level of activity generally have a smaller value of rotational velocity according to the activity--rotation relationship of stars (e.g., \citealt{1984ApJ...279..763N, 2011ApJ...743...48W, 2017A&A...600A..13A, 2017ApJ...834...85N, 2020ApJ...902..114W, 2023ApJS..264...12H}),
which can effectively mitigate the large $\Delta I_{\mathrm{H}{\alpha}}$ in panels g--i of Figure \ref{fig:I_index_increment_vs_vsini}.
If $v{\sin}i$ is less than 20\,km/s, the magnitude of $\Delta  I_{\mathrm{H}{\alpha}}$ in panels g--i of
Figure \ref{fig:I_index_increment_vs_vsini} would be below 0.05.
Putting the above discussions together, for the MRS spectral sample analyzed in this work, the absolute magnitude of $\Delta I_{\mathrm{H}{\alpha}}$ as a whole can be considered to be generally less than about 0.05,
which is on the same order of magnitude as the stochastic uncertainty of $I_{\mathrm{H}{\alpha}}$.

\subsection{$\mathrm{H}\alpha$ $R$-index} \label{sec:R_index}

The values of the $I_{\mathrm{H}{\alpha}}$ index obtained in Section \ref{sec:I_index} are affected by the continuum flux which in turn is a function of stellar atmospheric parameters.
For that reason, we employ the H$\alpha$ $R$-index (denoted by $R_{\mathrm{H}{\alpha}}$) as the secondary index of stellar H$\alpha$ chromospheric activity,
which is defined as the ratio of the stellar H$\alpha$ luminosity to the stellar bolometric luminosity (e.g, \citealt{2004PASP..116.1105W, 2011AJ....141...97W}), i.e.,
\begin{equation} \label{equ:R_index_definition}
  R_{\mathrm{H}{\alpha}} =
  \frac{L_{\mathrm{H}{\alpha}}}{L_\mathrm{bol}} =
  \frac{\mathscr{F}_{\mathrm{H}{\alpha}}}{{\mathscr{F}}_\mathrm{bol}},
\end{equation}
where $L$ represents the stellar luminosity and $\mathscr{F}$ represents the flux density (energy per unit time per unit area) on the stellar surface.
Note that $L=\mathscr{F} \cdot A_{\star}$, where $A_{\star}$ is the stellar surface area.

The $R_{\mathrm{H}{\alpha}}$ index defined in Equation (\ref{equ:R_index_definition}) eliminates the effect of continuum flux variation,
and therefore is more suitable for comparing the activity properties of different types of stars.

The $\mathscr{F}_\mathrm{bol}$ in Equation (\ref{equ:R_index_definition}) can be expressed by $T_\mathrm{eff}$ as
\begin{equation} \label{equ:bolometric_flux_density}
  \mathscr{F}_\mathrm{bol} = \sigma T_\mathrm{eff}^4,
\end{equation}
where $\sigma$ is the Stefan-Boltzmann constant.
The $\mathscr{F}_{\mathrm{H}{\alpha}}$ in Equation (\ref{equ:R_index_definition}) can be expressed as
\begin{equation} \label{equ:halpha_flux_density}
  \mathscr{F}_{\mathrm{H}{\alpha}} = 
  \Delta\lambda \cdot \overline{f}_{\mathrm{H}{\alpha}} =
  \Delta\lambda \cdot \overline{f}_\mathrm{cont} \cdot
  \frac{\overline{f}_{\mathrm{H}{\alpha}}}
  {\overline{f}_\mathrm{cont}} =
  \Delta\lambda \cdot \overline{f}_\mathrm{cont} \cdot
  \frac{\overline{F}_{\mathrm{H}{\alpha}}}{\overline{F}_\mathrm{cont}}=
  \Delta\lambda \cdot \overline{f}_\mathrm{cont} \cdot
  I_{\mathrm{H}{\alpha}},
\end{equation}
where $\Delta\lambda$ is the width of the center band of H$\alpha$ line 
(equaling to 0.25\,{\AA} as defined in Section \ref{sec:I_index}), 
$f$ represents the spectral flux density (energy per unit time per unit area per unit wavelength) on the stellar surface, 
$\overline{f}_{\mathrm{H}{\alpha}}$ is the mean value of $f$ in the center band of H$\alpha$ line, 
$\overline{f}_\mathrm{cont}$ is the mean value of $f$ in the two continuum bands defined in Section \ref{sec:I_index}, 
and $F$ represents the observed flux data of MRS which is the same as in Equation (\ref{equ:I_halpha_index_definition}).

By substituting Equations (\ref{equ:bolometric_flux_density}) and (\ref{equ:halpha_flux_density}) into Equation (\ref{equ:R_index_definition}), we obtain
\begin{equation} \label{equ:R_and_I_indices_relation}
  R_{\mathrm{H}{\alpha}} =
  \Delta\lambda \cdot 
  \frac{\overline{f}_\mathrm{cont}}{\sigma T_\mathrm{eff}^4} \cdot
  I_{\mathrm{H}{\alpha}}=
  \Delta\lambda \cdot 
  \chi \cdot
  I_{\mathrm{H}{\alpha}},
\end{equation}
where the factor 
\begin{equation} \label{equ:chi_definition}
  \chi = \overline{f}_\mathrm{cont}/(\sigma T_\mathrm{eff}^4)
\end{equation}
which is a function of stellar atmospheric parameters \citep{2004PASP..116.1105W} and in unit of {\AA}$^{-1}$.

To evaluate the value of $R_{\mathrm{H}{\alpha}}$ from $I_{\mathrm{H}{\alpha}}$ using Equation (\ref{equ:R_and_I_indices_relation}), one should first know the value of the factor $\chi$.
In this work, we obtain the $\chi$ values by using the library of synthetic spectra of stellar atmospheres presented in \citet{2013A&A...553A...6H}, 
which is based on the PHOENIX stellar atmospheric code \citep{1995ApJ...445..433A, 2016sf2a.conf..223A}.
The synthetic spectra of stellar atmospheres provide the data of $f$ with physical units,
thus the $\chi$ values can be calculated directly from the synthetic spectra by using Equation (\ref{equ:chi_definition}).
A comparison by \citet{2021A&A...649A..97L} between the synthetic spectra of \citet{2013A&A...553A...6H} and the observed spectra across the Hertzsprung-Russell diagram showed that a satisfactory representation of the continuum flux can be obtained by the synthetic spectra in the visual band with stellar effective temperature down to about 4000\,K (see \citealt{2021A&A...649A..97L} for details).
The library of stellar synthetic spectra by \citet{2013A&A...553A...6H} only provides the spectral data of a grid of stellar atmospheric parameters; 
for the stellar parameters between the grid points, the value of $\chi$ is obtained via interpolation based on the $\chi$ values on the grid.
The uncertainties of the obtained $\chi$ values (denoted by $\delta \chi$) depend on the uncertainties of the stellar atmospheric parameters ($\delta T_\mathrm{eff}$, $\delta\log\,g$, and $\delta\mathrm{[Fe/H]}$) of the MRS spectra,
and the latter has been provided by the {\tt LAMOST MRS Parameter Catalog}. 
The algorithm for the $\delta \chi$ estimation is described in Appendix \ref{sec:delta_chi_estimation}.

We calculate the values of $R_{\mathrm{H}{\alpha}}$ using Equation (\ref{equ:R_and_I_indices_relation}) for the selected MRS spectra of F-, G-, and K-type stars.
The stochastic uncertainties of the $R_{\mathrm{H}{\alpha}}$ values (denoted by $\delta R_{\mathrm{H}{\alpha}}$) are estimated from $\delta\chi$ and $\delta I_{\mathrm{H}{\alpha}}$ using the error propagation rules.
Five MRS spectra in the giant sample have $\log\,g$ values less than zero, which are out of the parameter scope of the synthetic spectra by \citet{2013A&A...553A...6H};
so the $\chi$ and $R_{\mathrm{H}{\alpha}}$ values of the five MRS spectra are not calculated.
In addition to the five spectra, 
the $\delta T_\mathrm{eff}$, $\delta\log\,g$, and $\delta\mathrm{[Fe/H]}$ values of a few MRS spectra are not fully available in the LAMOST catalog, 
and the values of $\log\,g - \delta\log\,g$ of very few spectra are out of the parameter scope of the synthetic spectra;
for these spectra (47,495 in total), the uncertainties of $\chi$, and hence the uncertainties of $R_{\mathrm{H}{\alpha}}$, are not estimated.

\begin{figure}
  \centering
  \includegraphics[width=1.02\textwidth]{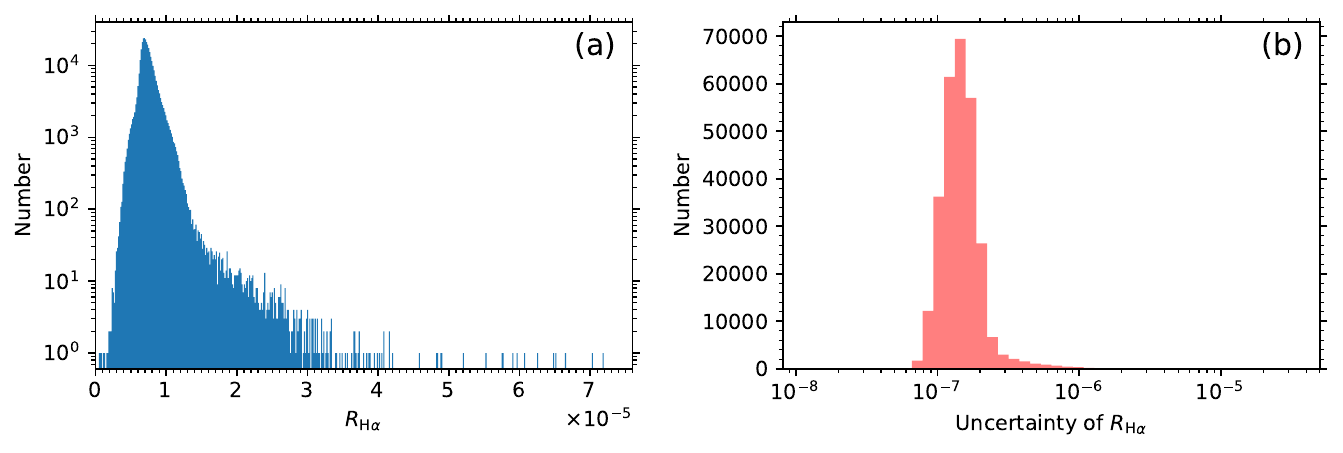}
  \caption{
    Histograms of the evaluated $R_{\mathrm{H}{\alpha}}$ index values (panel a) and their uncertainties (panel b) for the selected MRS spectra of F-, G-, and K-type stars.
  }
  \label{fig:histogram_R_halpha_and_err}
\end{figure}

Figure \ref{fig:histogram_R_halpha_and_err} shows the histograms of the evaluated $R_{\mathrm{H}{\alpha}}$ index values and the estimated uncertainties of $R_{\mathrm{H}{\alpha}}$.
It can be seen from Figure \ref{fig:histogram_R_halpha_and_err} that the $R_{\mathrm{H}{\alpha}}$ values are on the order of magnitude of $10^{-5}$,
and the uncertainties of $R_{\mathrm{H}{\alpha}}$ are on the order of magnitude of $10^{-7}$.
All data of the obtained $R_{\mathrm{H}{\alpha}}$ index values and their uncertainties, as well as the associated $\chi$ values and their uncertainties, are available in the online dataset of this paper (see Section \ref{sec:dataset}). 
The missing values of $R_{\mathrm{H}{\alpha}}$, $\delta R_{\mathrm{H}{\alpha}}$, $\chi$, and $\delta\chi$ mentioned above are indicated by $-9999.0$ in the dataset.

\section{Distribution of the $\mathrm{H}\alpha$ chromospheric activity} \label{sec:results}

By using the $R_{\mathrm{H}{\alpha}}$ index values obtained in Section \ref{sec:halpha_indices} for the selected MRS spectral sample of F-, G-, and K-type stars,
we analyze in detail the distribution of H$\alpha$ chromospheric activity with respect to stellar atmospheric parameters and main-sequence/giant categories.
In Section \ref{sec:R_index_with_teff_logg_feh}, we analyze and compare the $R_{\mathrm{H}{\alpha}}$ index distributions with individual stellar atmospheric parameters ($T_\mathrm{eff}$, $\log\,g$, and $\mathrm{[Fe/H]}$) for the main-sequence and giant samples.
In Sections \ref{sec:R_index_in_teff_logg_space} and \ref{sec:R_index_in_teff_feh_space}, we further investigate the distributions of the $R_{\mathrm{H}{\alpha}}$ index values in the $T_\mathrm{eff}$ -- $\log\,g$ and $T_\mathrm{eff}$ -- $\mathrm{[Fe/H]}$ parameter spaces, respectively.

\subsection{Distribution of the $R_{\mathrm{H}{\alpha}}$ index with $T_\mathrm{eff}$, $\log\,g$, and $\mathrm{[Fe/H]}$ for the main-sequence and giant samples} \label{sec:R_index_with_teff_logg_feh}

\begin{figure}
  \centering
  \includegraphics[width=1.095\textwidth]{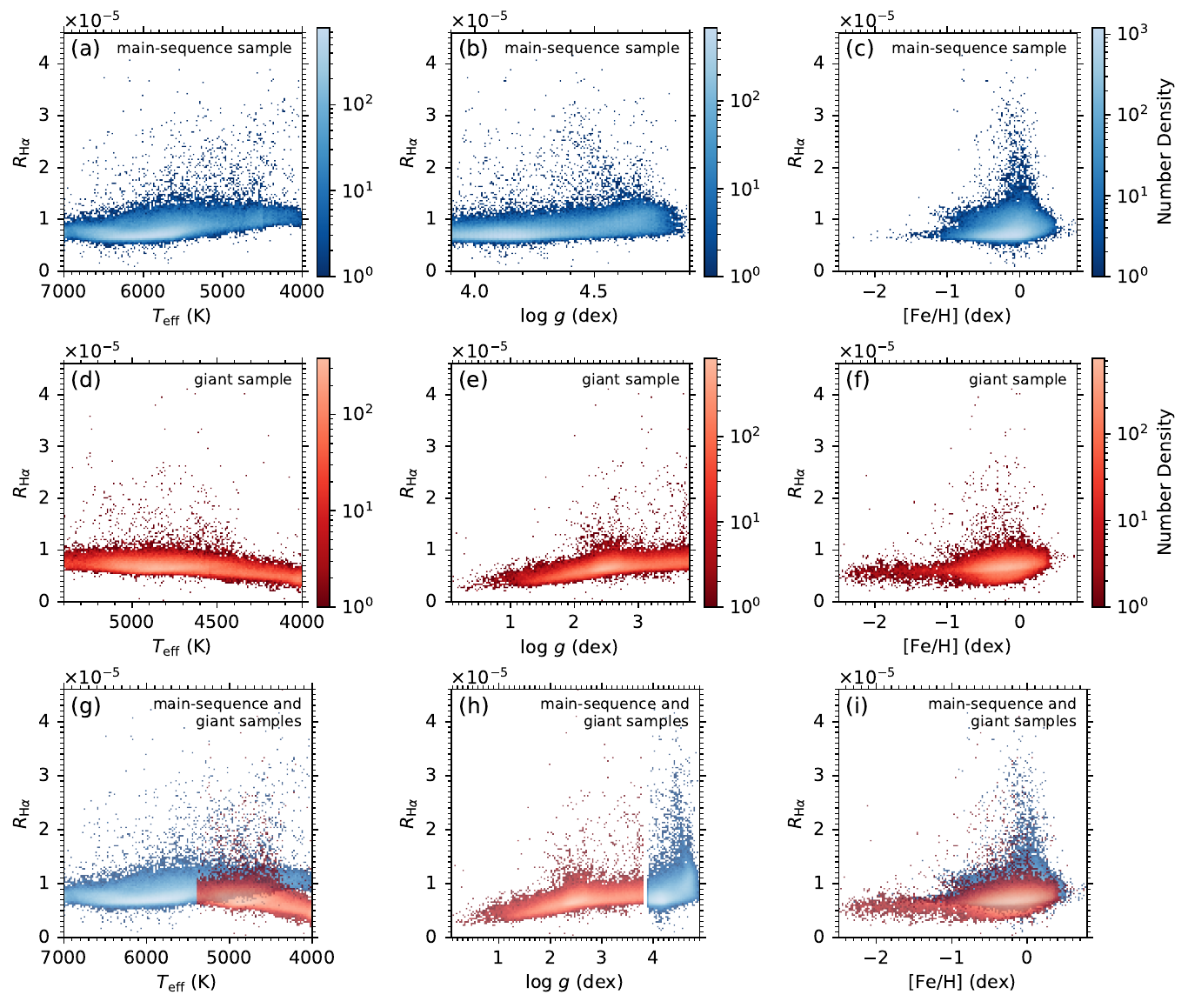}
  \caption{Distribution of the $R_{\mathrm{H}{\alpha}}$ index values with $T_\mathrm{eff}$ (left column), $\log\,g$ (middle column), and $\mathrm{[Fe/H]}$ (right column) for the selected MRS spectra of main-sequence sample (top row), giant sample (middle row), and both categories (bottom row).
  Color scale indicates number density of the spectral sample (main-sequence sample in blue and giant sample in red).}
  \label{fig:R_index_with_teff_logg_feh}
\end{figure}

Figure \ref{fig:R_index_with_teff_logg_feh} shows the distribution of the $R_{\mathrm{H}{\alpha}}$ index values with effective temperature $T_\mathrm{eff}$ (left column), surface gravity $\log\,g$ (middle column), and metallicity $\mathrm{[Fe/H]}$ (right column) for the selected MRS spectra of main-sequence sample (top row), giant sample (middle row), and both categories (bottom row).

The distributions of $R_{\mathrm{H}{\alpha}}$ with $T_\mathrm{eff}$ for the main-sequence and giant samples are distinctly different as exhibited in the left column (panels a, d, and g) of Figure \ref{fig:R_index_with_teff_logg_feh}.

As shown in Figure \ref{fig:R_index_with_teff_logg_feh}a,
there are three apparent envelopes in the $R_{\mathrm{H}{\alpha}}$ versus $T_\mathrm{eff}$ distribution plot of the main-sequence sample:  
a bowl-shaped lower envelope, a hill-shaped middle envelope, and an upper envelope continuing to increase from hotter to cooler stars. 
These envelopes are characterized by a sharp change in the number density of spectral sample.
The lower envelope in Figure \ref{fig:R_index_with_teff_logg_feh}a has a minimum $R_{\mathrm{H}{\alpha}}$ value of about $0.55 \times 10^{-5}$ at about 6200\,K,
and the middle envelope has a maximum $R_{\mathrm{H}{\alpha}}$ value of about $1.55 \times 10^{-5}$ at about 5600\,K.
The distinction between the upper envelope and the middle envelope in Figure \ref{fig:R_index_with_teff_logg_feh}a is most pronounced in the $T_\mathrm{eff}$ range cooler than 6000\,K, 
and the $R_{\mathrm{H}{\alpha}}$ values of the upper envelope continue to increase from the middle envelope value at about 6000\,K to about $3.70 \times 10^{-5}$ at 4000\,K.

For the giant sample, the $R_{\mathrm{H}{\alpha}}$ versus $T_\mathrm{eff}$ distribution plot (see Figure \ref{fig:R_index_with_teff_logg_feh}d) also presents the lower, middle, and upper envelopes.
However, the lower envelope of the giant sample in Figure \ref{fig:R_index_with_teff_logg_feh}d continues to decease from hotter to cooler stars, 
and both the middle and upper envelopes first increase with decrease of $T_\mathrm{eff}$ and then drop to a lower activity level at about 4300\,K.
The middle envelope in Figure \ref{fig:R_index_with_teff_logg_feh}d reaches a maximum $R_{\mathrm{H}{\alpha}}$ value of about $1.55 \times 10^{-5}$ at about 4550\,K,
and the upper envelope reaches a maximum $R_{\mathrm{H}{\alpha}}$ value of about $2.70 \times 10^{-5}$ at about 4800\,K. 
These features in $R_{\mathrm{H}{\alpha}}$ versus $T_\mathrm{eff}$ distribution plot of the giant sample are distinctly different from those of the main-sequence sample, revealing the different activity characteristics at different stages of stellar evolution.

The middle column of Figure \ref{fig:R_index_with_teff_logg_feh} (panels b, e, and h) shows the distributions of $R_{\mathrm{H}{\alpha}}$ with $\log\,g$ for the main-sequence and giant samples.
The distribution plot in Figure \ref{fig:R_index_with_teff_logg_feh}h ($R_{\mathrm{H}{\alpha}}$ versus $\log\,g$ for both the main-sequence and giant samples) suggests that a larger surface gravity favors a higher $R_{\mathrm{H}{\alpha}}$ index,
but the shape of the envelopes in the plot is complicated and the trend is not monotonic.

From the right column of Figure \ref{fig:R_index_with_teff_logg_feh} (panels c, f, and i) it can be seen that the distributions of $R_{\mathrm{H}{\alpha}}$ with $\mathrm{[Fe/H]}$ for the main-sequence and giant samples are similar in the sense that the upper envelopes in the $R_{\mathrm{H}{\alpha}}$ versus $\mathrm{[Fe/H]}$ distribution plots are higher for stars with $\mathrm{[Fe/H]}$ greater than about $-1.0$,
and the metal-poor stars with $\mathrm{[Fe/H]}$ less than about $-1.0$ generally have a lower level of H$\alpha$ activity.
The fact that the lowest-metallicity stars hardly exhibit high H$\alpha$ indices implies that they might be very old stars (e.g., \citealt{2010ARA&A..48..581S, 2015ARA&A..53..631F}).

The bottom row of Figure \ref{fig:R_index_with_teff_logg_feh} (panels g, h, and i) displays the distributions of $R_{\mathrm{H}{\alpha}}$ with the three stellar atmospheric parameters for both the main-sequence and giant samples (main-sequence sample in blue and giant sample in red)
so that the $R_{\mathrm{H}{\alpha}}$ index distributions of the two stellar categories can be compared directly.
The distribution plots of $R_{\mathrm{H}{\alpha}}$ versus the three stellar parameters all exhibit that the H$\alpha$ chromospheric activity of giant stars is on average lower than that of main-sequence stars, which is consistent with the results in the literature (e.g., \citealt{1991A&A...251..199P, 2005A&A...431..329L}).

It needs to be mentioned that some spectra have isolated ultra-high $R_{\mathrm{H}{\alpha}}$ values relative to the upper envelope of the $R_{\mathrm{H}{\alpha}}$ distributions as displayed in Figure \ref{fig:R_index_with_teff_logg_feh}.
These ultra-high values of H$\alpha$ indices can also be seen in the histogram of $R_{\mathrm{H}{\alpha}}$ in Figure \ref{fig:histogram_R_halpha_and_err}a as well as in the histogram of $I_{\mathrm{H}{\alpha}}$ in Figure \ref{fig:histogram_I_halpha_and_err}a.
Most of the ultra-high values of H$\alpha$ indices come from the transient eruption events (such as stellar flares) as indicated by the associated MRS single-exposure spectra, which show rapid time-sequence evolution of H$\alpha$ line profiles within a couple of hours (e.g., \citealt{2022ApJ...928..180W, 2022A&A...663A.140L}).
In this work, we focus on the stellar chromospheric activity in the usual sense,
i.e., the activity associated with the steady state of stellar chromosphere (\citealt{2008LRSP....5....2H}).
The spectra of the transient eruption events will be studied in the future research.

We also examine those few data points beneath the lower envelope of the $R_{\mathrm{H}{\alpha}}$ distributions shown in Figure \ref{fig:R_index_with_teff_logg_feh}.
Most of the data points are associated with abnormal spectral profiles in original MRS spectral data and can therefore be regarded as outliers.

\subsection{Distribution of the $R_{\mathrm{H}{\alpha}}$ index in the $T_\mathrm{eff}$ -- $\log\,g$ parameter space} \label{sec:R_index_in_teff_logg_space}

To investigate the connections between the $R_{\mathrm{H}{\alpha}}$ index distributions with respect to individual stellar atmospheric parameters presented in Section \ref{sec:R_index_with_teff_logg_feh},
in this subsection we further analyze the distribution of the $R_{\mathrm{H}{\alpha}}$ values in the $T_\mathrm{eff}$ -- $\log\,g$ parameter space,
and the $R_{\mathrm{H}{\alpha}}$ distribution in the $T_\mathrm{eff}$ -- $\mathrm{[Fe/H]}$ parameter space will be analyzed in Section \ref{sec:R_index_in_teff_feh_space}.

\begin{figure}
  \centering
  \includegraphics[width=0.90\textwidth]{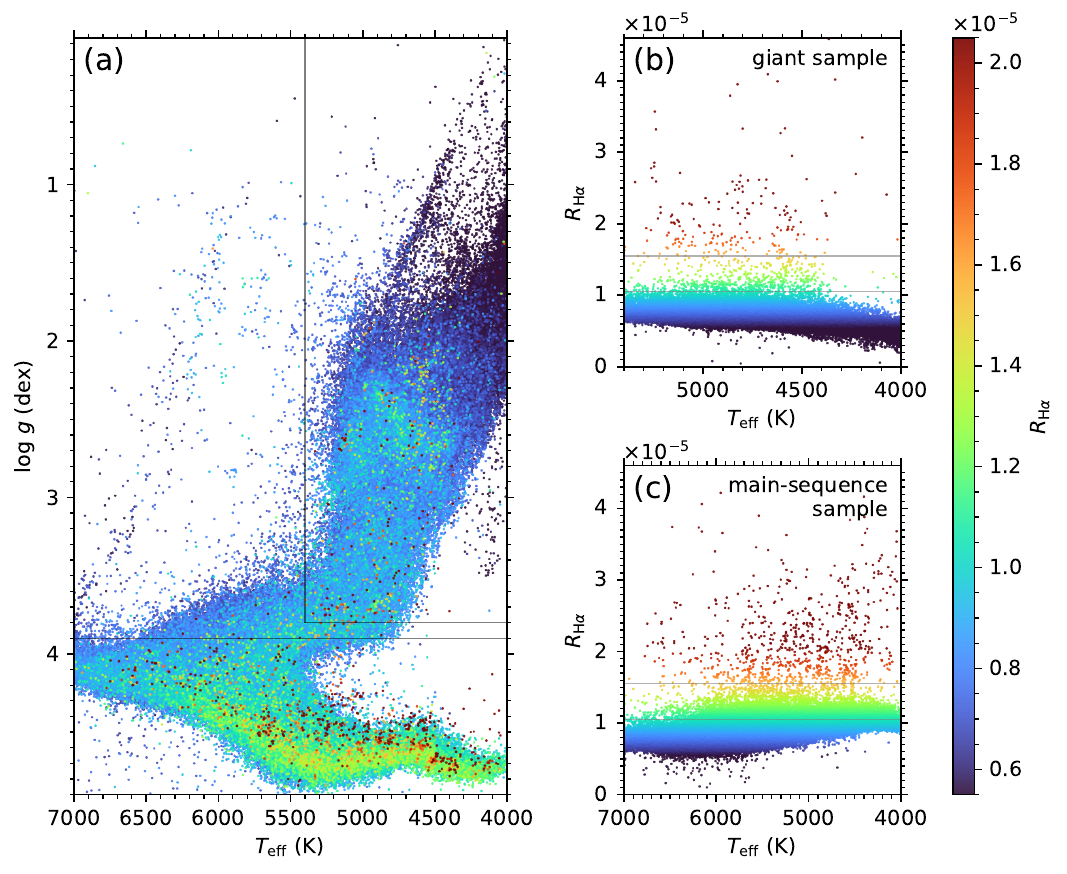}
  \caption{
  (a) Distribution of the $R_{\mathrm{H}{\alpha}}$ index values in the $T_\mathrm{eff}$ -- $\log\,g$ parameter space for the selected MRS spectra of F-, G-, and K-type stars.
  The value of $R_{\mathrm{H}{\alpha}}$ is indicated by color scale.
  The data points in the diagram are stacked in ascending order of their $R_{\mathrm{H}{\alpha}}$ values, with the maximum value at the highest layer and the minimum value at the lowest layer.
  The horizontal and vertical lines in the diagram are the dividing lines for the main-sequence and giant samples (same as in Figure \ref{fig:number_density_teff_vs_logg}).
  (b) Scatter plot of $R_{\mathrm{H}{\alpha}}$ versus $T_\mathrm{eff}$ for the giant sample.
  (c) Scatter plot of $R_{\mathrm{H}{\alpha}}$ versus $T_\mathrm{eff}$ for the main-sequence sample.
  Panels b and c use the same $R_{\mathrm{H}{\alpha}}$ color scale as panel a.
  The two horizontal lines in panels b and c represent $R_{\mathrm{H}{\alpha}}=1.05 \times 10^{-5}$ and $1.55 \times 10^{-5}$, respectively,
  for distinguishing the three levels of $R_{\mathrm{H}{\alpha}}$ values (see main text for details).
  }
  \label{fig:R_index_in_teff_logg_space}
\end{figure}

Figure \ref{fig:R_index_in_teff_logg_space}a shows the distribution of the $R_{\mathrm{H}{\alpha}}$ index values in the $T_\mathrm{eff}$ -- $\log\,g$ parameter space for the selected MRS spectra of F-, G-, and K-type stars.
The value of $R_{\mathrm{H}{\alpha}}$ is indicated by color scale.
The data points in Figure \ref{fig:R_index_in_teff_logg_space}a are stacked in ascending order of their $R_{\mathrm{H}{\alpha}}$ values, with the maximum value at the highest layer and the minimum value at the lowest layer.
The horizontal and vertical lines in Figure \ref{fig:R_index_in_teff_logg_space}a are the dividing lines for the main-sequence and giant samples (same as in Figure \ref{fig:number_density_teff_vs_logg}).
For comparison with the results in Section \ref{sec:R_index_with_teff_logg_feh},
we also show the scatter plots of $R_{\mathrm{H}{\alpha}}$ versus $T_{\rm  eff}$ for the giant sample and the main-sequence sample in Figures \ref{fig:R_index_in_teff_logg_space}b and \ref{fig:R_index_in_teff_logg_space}c, respectively,
with the same $R_{\mathrm{H}{\alpha}}$ color scale as in Figure \ref{fig:R_index_in_teff_logg_space}a.

As demonstrated in Figure \ref{fig:R_index_in_teff_logg_space}, the magnitude of $R_{\mathrm{H}{\alpha}}$ can be roughly divided into three levels:
the lower level ($R_{\mathrm{H}{\alpha}}$ less than about $1.05 \times 10^{-5}$), the middle level ($R_{\mathrm{H}{\alpha}}$ between about $1.05 \times 10^{-5}$ and $1.55 \times 10^{-5}$), and the upper level ($R_{\mathrm{H}{\alpha}}$ greater than about $1.55 \times 10^{-5}$),
which are displayed in blue, green, and red, respectively, in Figure \ref{fig:R_index_in_teff_logg_space}.
The three levels of $R_{\mathrm{H}{\alpha}}$ values are also distinguished by two horizontal lines (representing $R_{\mathrm{H}{\alpha}}=1.05 \times 10^{-5}$ and $1.55 \times 10^{-5}$, respectively) in Figures \ref{fig:R_index_in_teff_logg_space}b and \ref{fig:R_index_in_teff_logg_space}c.
It can be seen from Figures \ref{fig:R_index_in_teff_logg_space}b and \ref{fig:R_index_in_teff_logg_space}c that the lower, middle, and upper envelopes in the $R_{\mathrm{H}{\alpha}}$ versus $T_\mathrm{eff}$ distribution plots discussed in Section \ref{sec:R_index_with_teff_logg_feh} are roughly associated with the lower, middle, and upper levels of $R_{\mathrm{H}{\alpha}}$ values, respectively.

The distribution areas of the three levels of $R_{\mathrm{H}{\alpha}}$ values in the $T_\mathrm{eff}$ -- $\log\,g$ parameter space are different as exhibited in Figure \ref{fig:R_index_in_teff_logg_space}a.
The lower-level $R_{\mathrm{H}{\alpha}}$ values (displayed in blue) occupy the largest area in the $T_\mathrm{eff}$ -- $\log\,g$ space and encompass the vast majority of the spectral sample.
The distribution area of the middle-level $R_{\mathrm{H}{\alpha}}$ values (displayed in green) is less than that of the of the lower-level $R_{\mathrm{H}{\alpha}}$ values.
For the main-sequence sample, most of the middle-level $R_{\mathrm{H}{\alpha}}$ values distribute along the usual main-sequence strip zone in the $T_\mathrm{eff}$ -- $\log\,g$ space;
while for the giant sample, the middle-level $R_{\mathrm{H}{\alpha}}$ values spread over an irregularly shaped area.
The upper-level $R_{\mathrm{H}{\alpha}}$ values (displayed in red) occupy the smallest area in the $T_\mathrm{eff}$ -- $\log\,g$ space,
and the location of the upper-level $R_{\mathrm{H}{\alpha}}$ values in the $T_\mathrm{eff}$ -- $\log\,g$ space does not exactly coincide with that of the middle-level $R_{\mathrm{H}{\alpha}}$ values for both the main-sequence and giant samples.
For the main-sequence sample, the difference between the locations of the middle-level and upper-level $R_{\mathrm{H}{\alpha}}$ values in the $T_\mathrm{eff}$ -- $\log\,g$ space suggests that they might be associated with stars at different evolution stages and ages (e.g., \citealt{2013ApJS..208....9P, 2015A&A...577A..42B}),
that is, the younger the star, the higher the activity level.

\begin{figure}
  \centering
  \includegraphics[width=0.55\textwidth]{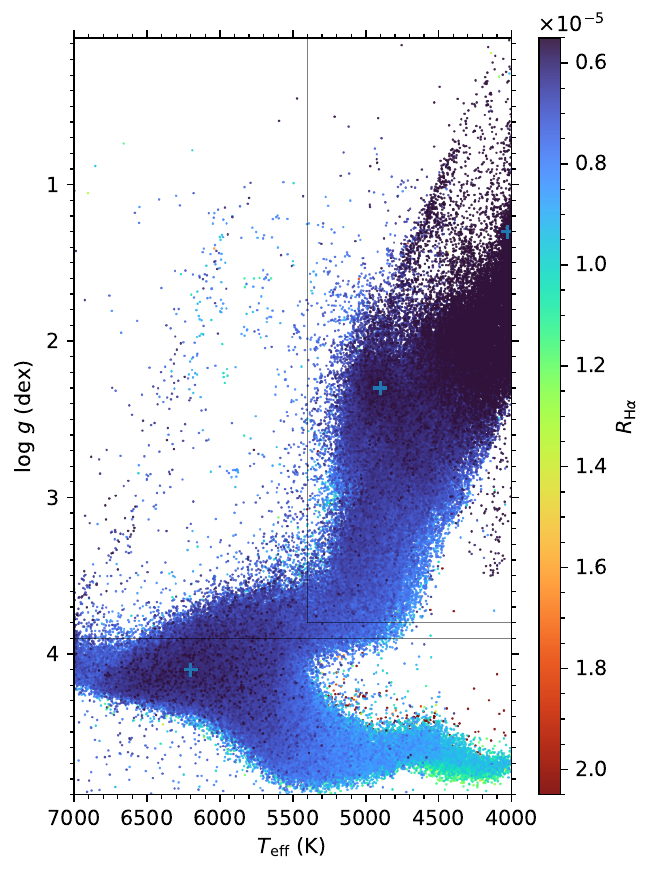}
  \caption{
  Distribution of the bottommost $R_{\mathrm{H}{\alpha}}$ index values in the $T_\mathrm{eff}$ -- $\log\,g$ parameter space for the selected MRS spectra of F-, G-, and K- type stars.
  The value of $R_{\mathrm{H}{\alpha}}$ is indicated by color scale (same as in Figure \ref{fig:R_index_in_teff_logg_space}, but with an inverted $R_{\mathrm{H}{\alpha}}$-axis of colorbar).
  The data points in the diagram are stacked in reverse order of their $R_{\mathrm{H}{\alpha}}$ values, with the minimum value at the highest layer and the maximum value at the lowest layer.
  The horizontal and vertical lines in the diagram are the dividing lines for the main-sequence and giant samples (same as in Figure \ref{fig:number_density_teff_vs_logg}).
  The `+' symbol indicates the local areas with the minimum $R_{\mathrm{H}{\alpha}}$ values (displayed in deep blue; see main text for details).
  }
  \label{fig:R_index_in_teff_logg_space_bottommost}
\end{figure}

In Figure \ref{fig:R_index_in_teff_logg_space}a, the bottommost data points of the $R_{\mathrm{H}{\alpha}}$ values are obscured by the higher data points and hence are not visible.
In Figure \ref{fig:R_index_in_teff_logg_space_bottommost} we show the dedicated diagram of the $R_{\mathrm{H}{\alpha}}$ index distribution in the $T_\mathrm{eff}$ -- $\log\,g$ parameter space for the bottommost $R_{\mathrm{H}{\alpha}}$ values.
The color scale of $R_{\mathrm{H}{\alpha}}$ in Figure \ref{fig:R_index_in_teff_logg_space_bottommost} is the same as in Figure \ref{fig:R_index_in_teff_logg_space},
but the data points are stacked in reverse order of their $R_{\mathrm{H}{\alpha}}$ values,
i.e., the minimum value at the highest layer and the maximum value at the lowest layer.

As shown in Figures \ref{fig:R_index_in_teff_logg_space}b and \ref{fig:R_index_in_teff_logg_space}c, 
the minimum $R_{\mathrm{H}{\alpha}}$ values (displayed in deep blue) are around $0.55 \times 10^{-5}$ for the main-sequence sample and less than about $0.60 \times 10^{-5}$ for the giant sample.
These minimum $R_{\mathrm{H}{\alpha}}$ values are distributed in separate deep blue areas in the $T_\mathrm{eff}$ -- $\log\,g$ parameter space as demonstrated in Figure \ref{fig:R_index_in_teff_logg_space_bottommost}.
For the giant sample, there are two deep blue areas: one is around upper-right corner of the parameter space, and another is around $T_\mathrm{eff}=4900\,{\rm K}$ and $\log\,g=2.3$.
For the main-sequence sample, there is one deep blue area which is around $T_\mathrm{eff}=6200\,{\rm K}$ and $\log\,g=4.1$.
The location of these deep blue areas with minimum $R_{\mathrm{H}{\alpha}}$ values is marked with a `+' symbol in Figure \ref{fig:R_index_in_teff_logg_space_bottommost}.
The distribution of the minimum $R_{\mathrm{H}{\alpha}}$ areas in Figure \ref{fig:R_index_in_teff_logg_space_bottommost} suggests that,
besides the parameter $T_\mathrm{eff}$, other stellar atmospheric parameters such as $\log\,g$ have equal significance in determining the bottom envelope of $R_{\mathrm{H}{\alpha}}$ distribution in the parameter space.

\subsection{Distribution of the $R_{\mathrm{H}{\alpha}}$ index in the $T_\mathrm{eff}$ -- $\mathrm{[Fe/H]}$ parameter space for the main-sequence and giant samples} \label{sec:R_index_in_teff_feh_space}

\begin{figure}
  \centering
  \includegraphics[width=1.121\textwidth]{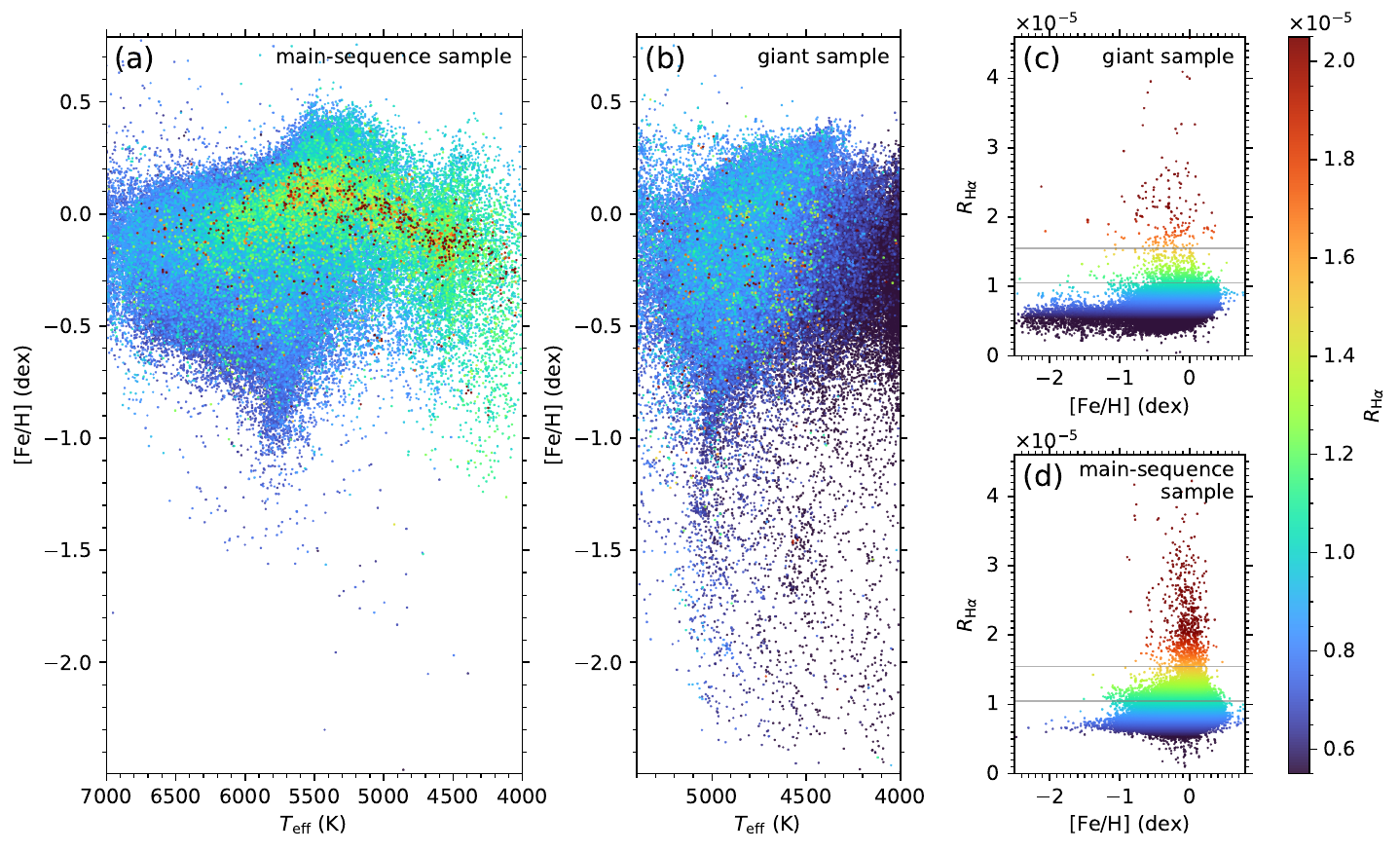}
  \caption{
    (a, b) Distributions of the $R_{\mathrm{H}{\alpha}}$ index values in the $T_\mathrm{eff}$ -- $\mathrm{[Fe/H]}$ parameter space for the selected MRS spectra of the main-sequence sample (panel a) and the giant sample (panel b).
   The value of $R_{\mathrm{H}{\alpha}}$ is indicated by color scale (same as in Figure \ref{fig:R_index_in_teff_logg_space}).
   The data points in the diagrams are stacked in ascending order of their $R_{\mathrm{H}{\alpha}}$ values, with the maximum value at the highest layer and the minimum value at the lowest layer.
   (c, d) Scatter plots of $R_{\mathrm{H}{\alpha}}$ versus $\mathrm{[Fe/H]}$ for the giant sample (panel c) and the main-sequence sample (panel d).
   Panels c and d use the same $R_{\mathrm{H}{\alpha}}$ color scale as panels a and b.
   The two horizontal lines in panels c and d represent $R_{\mathrm{H}{\alpha}}=1.05 \times 10^{-5}$ and $1.55 \times 10^{-5}$, respectively,
  for distinguishing the three levels of $R_{\mathrm{H}{\alpha}}$ values (see main text for details).
  }
  \label{fig:R_index_in_teff_feh_space}
\end{figure}

Figures \ref{fig:R_index_in_teff_feh_space}a and \ref{fig:R_index_in_teff_feh_space}b show the distributions of the $R_{\mathrm{H}{\alpha}}$ index values in the $T_\mathrm{eff}$ -- $\mathrm{[Fe/H]}$ parameter space for the selected MRS spectra of the main-sequence sample and the giant sample, respectively,
with the same $R_{\mathrm{H}{\alpha}}$ color scale as in Figure \ref{fig:R_index_in_teff_logg_space}.
The data points are stacked in ascending order of their $R_{\mathrm{H}{\alpha}}$ values, with the maximum value at the highest layer and the minimum value at the lowest layer.
For comparison with the results in Section \ref{sec:R_index_with_teff_logg_feh},
we also show the scatter plots of $R_{\mathrm{H}{\alpha}}$ versus $\mathrm{[Fe/H]}$ for the giant sample and the main-sequence sample in Figures \ref{fig:R_index_in_teff_feh_space}c and \ref{fig:R_index_in_teff_feh_space}d, respectively.
The two horizontal lines in Figures \ref{fig:R_index_in_teff_feh_space}c and \ref{fig:R_index_in_teff_feh_space}d represent $R_{\mathrm{H}{\alpha}}=1.05 \times 10^{-5}$ and $1.55 \times 10^{-5}$, respectively,
for distinguishing the three levels of $R_{\mathrm{H}{\alpha}}$ values described in Section \ref{sec:R_index_in_teff_logg_space}.

It can be seen from Figure \ref{fig:R_index_in_teff_feh_space} that, for the main-sequence sample,
the middle-level $R_{\mathrm{H}{\alpha}}$ values (displayed in green) are mainly distributed in a $\Lambda$-shaped strip area in the $T_\mathrm{eff}$ -- $\mathrm{[Fe/H]}$ space, which swings up and down around $\mathrm{[Fe/H]}=0.0$ (solar metallicity) with the vertex at about $T_\mathrm{eff} \sim$ 5400\,K;
while for the giant sample, the middle-level $R_{\mathrm{H}{\alpha}}$ values spread over an irregularly shaped area in the $T_\mathrm{eff}$ -- $\mathrm{[Fe/H]}$ space,
which is distinctly different from that of the main-sequence sample.

Figure \ref{fig:R_index_in_teff_feh_space} also exhibits that the distribution area of the upper-level $R_{\mathrm{H}{\alpha}}$ values (displayed in red) basically coincides with the distribution area of the middle-level $R_{\mathrm{H}{\alpha}}$ values in the $T_\mathrm{eff}$ -- $\mathrm{[Fe/H]}$ parameter space for both the main-sequence and giant samples,
which results in a spike pattern of the upper-level $R_{\mathrm{H}{\alpha}}$ values in Figure \ref{fig:R_index_in_teff_feh_space}d for the main-sequence sample and a fan-shaped pattern of the upper-level $R_{\mathrm{H}{\alpha}}$ values in Figure \ref{fig:R_index_in_teff_feh_space}c for the giant sample.
The coincidence of the locations of the middle-level and upper-level $R_{\mathrm{H}{\alpha}}$ values in the $T_\mathrm{eff}$ -- $\mathrm{[Fe/H]}$ space suggests that they might have the same origin on stars, though corresponding to different levels of activity.

\section{Dataset of $\mathrm{H}\alpha$ activity indices} \label{sec:dataset}

The $I_{\mathrm{H}{\alpha}}$ index data and the $R_{\mathrm{H}{\alpha}}$ index data of the LAMOST MRS spectra of F-, G-, and K-type stars analyzed in this work, 
as well as the related spectroscopic parameters from the LAMOST catalogs, 
are stored in a CSV-format file (file name: {\tt\string Halpha\_activity\_indices\_LAMOST\_MRS\_DR8.csv}) which is available in the online dataset of this paper.
The web link of the dataset is \url{https://doi.org/10.5281/zenodo.7396963}.
The columns included in the dataset are listed and briefly described in Table \ref{tab:dataset_columns}.
Some H$\alpha$ indices and spectroscopic parameters are not available for a few of MRS spectra, which are indicated by a value of $-9999.0$ in the dataset.

\begin{table}[tbp]
\centering
\caption{Columns included in the dataset of this paper.}
\label{tab:dataset_columns}
\begin{tabular*}{\textwidth}{@{\extracolsep\fill}lcp{8.69cm}}
\toprule
Column Name & Unit           & Description \\
            & (if available) & \\
\midrule
{\tt\string obsid} &  & unique observation identifier of LAMOST MRS \\
{\tt\string ra} & degree & right ascension of the observed object\\
{\tt\string dec} & degree & declination of the observed object \\
{\tt\string obsdate}  &  &  UTC date of the observation (format: YYYY-MM-DD)  \\
{\tt\string fitsname} &  & MRS FITS data file name \\
{\tt\string sn\_B} &  & signal-to-noise ratio of the blue band of MRS (${\rm S/N}_B $) \\
{\tt\string sn\_R} &  & signal-to-noise ratio of the red band of MRS (${\rm S/N}_R $) \\
{\tt\string band\_para} &  & indicating which band of MRS is used by the LASP 
to determine the stellar atmospheric parameters; `{\tt\string B}' representing the blue band and `{\tt\string R}' representing the red band \\
{\tt\string teff} & K & effective temperature ($T_\mathrm{eff}$) determined by the LASP \\
{\tt\string teff\_err} & K & uncertainty of $T_\mathrm{eff}$ \\
{\tt\string logg} & dex & surface gravity ($\log\,g$) determined by the LASP; $g$ in unit of ${\rm cm}/{{\rm s}^{2}}$ \\
{\tt\string logg\_err} & dex & uncertainty of $\log\,g$ \\
{\tt\string feh} & dex & metallicity ($\mathrm{[Fe/H]}$) determined by the LASP \\
{\tt\string feh\_err} & dex & uncertainty of $\mathrm{[Fe/H]}$ \\
{\tt\string rv\_r0} & km/s & radial velocity determined based on the red band data of MRS \\
{\tt\string rv\_r0\_err} & km/s & uncertainty of {\tt\string rv\_r0} \\
{\tt\string category}* &  & `{\tt\string main-sequence}' or `{\tt\string giant}' or `{\tt\string intermediate-zone}', as defined in Section \ref{sec:spectral_sample_selection} \\
{\tt\string I\_halpha}* &  & $I_{\mathrm{H}{\alpha}}$ index, as defined in Section \ref{sec:I_index} \\
{\tt\string I\_halpha\_err}* &  & uncertainty of $I_{\mathrm{H}{\alpha}}$ \\
{\tt\string chi}* & {\AA}$^{-1}$ & $\chi$ factor, as defined in Section \ref{sec:R_index} \\
{\tt\string chi\_err}* & {\AA}$^{-1}$ & uncertainty of $\chi$ \\
{\tt\string R\_halpha}* &  & $R_{\mathrm{H}{\alpha}}$ index, as defined in Section \ref{sec:R_index} \\
{\tt\string R\_halpha\_err}* &  & uncertainty of $R_{\mathrm{H}{\alpha}}$ \\
\botrule
\end{tabular*}
\footnotetext{Note: The columns marked with an asterisk are provided by this work. Other columns are from the MRS catalogs of LAMOST DR8 v1.1.}
\end{table}

\section{Conclusion and discussion} \label{sec:conclusion}

In this paper, we investigated the distribution of the stellar H$\alpha$ chromospheric activity with respect to stellar atmospheric parameters ($T_\mathrm{eff}$, $\log\,g$, and $\mathrm{[Fe/H]}$) and main-sequence/giant categories for the F-, G-, and K-type stars observed by the LAMOST MRS.
A total of 329,294 MRS spectra from LAMOST DR8 were used in the analysis.
The H$\alpha$ activity index ($I_{\mathrm{H}{\alpha}}$) and the H$\alpha$ $R$-index ($R_{\mathrm{H}{\alpha}}$) are evaluated for the MRS spectra.
The H$\alpha$ chromospheric activity distributions of the main-sequence and giant samples with respect to individual stellar parameters, as well as in the $T_\mathrm{eff}$ -- $\log\,g$ and $T_\mathrm{eff}$ -- $\mathrm{[Fe/H]}$ parameter spaces, were analyzed based on the $R_{\mathrm{H}{\alpha}}$ index.

We have found certain trends of the $R_{\mathrm{H}{\alpha}}$ index distribution with stellar effective temperature ($T_\mathrm{eff}$),
which are distinctly different for the main-sequence and giant samples.
For the main-sequence sample, the distribution has a bowl-shaped lower envelope with a minimum at about 6200\,K, a hill-shaped middle envelope with a maximum at about 5600\,K, and an upper envelope continuing to increase from hotter to cooler stars;
while for the giant sample, the middle and upper envelopes of the $R_{\mathrm{H}{\alpha}}$ index distribution first increases with a decrease of $T_\mathrm{eff}$ and then drop to a lower activity level at about 4300\,K.
The difference in the $R_{\mathrm{H}{\alpha}}$ versus $T_\mathrm{eff}$ distributions of the main-sequence and giant samples reveals the different activity characteristics at different stages of stellar evolution.

The distribution of $R_{\mathrm{H}{\alpha}}$ with $\log\,g$ for the main-sequence and giant samples suggests that a larger surface gravity favors a higher $R_{\mathrm{H}{\alpha}}$ index, 
but the shape of the envelopes of the $R_{\mathrm{H}{\alpha}}$ distribution with $\log\,g$ is complicated and the trend is not monotonic.

Although the morphology of the $R_{\mathrm{H}{\alpha}}$ index distribution with stellar metallicity ([Fe/H]) is different for the main-sequence and giant samples,
it is found that the upper envelope of the $R_{\mathrm{H}{\alpha}}$ distribution is higher for stars with $\mathrm{[Fe/H]}$ greater than about $-1.0$,
and the metal-poor stars with $\mathrm{[Fe/H]}$ less than about $-1.0$ generally have a lower level of H$\alpha$ activity.
This property is the same for both the main-sequence and giant samples.
The fact that the lowest-metallicity stars hardly exhibit high H$\alpha$ indices implies that they might be very old stars.

The distributions of the $R_{\mathrm{H}{\alpha}}$ index with the three stellar atmospheric parameters for the main-sequence and giant samples exhibit that the H$\alpha$ chromospheric activity of giant stars is on average lower than that of main-sequence stars, which is consistent with the results in the literature.

In the $T_\mathrm{eff}$ -- $\log\,g$ and $T_\mathrm{eff}$ -- $\mathrm{[Fe/H]}$ stellar parameter spaces,
it is found that the lower-level $R_{\mathrm{H}{\alpha}}$ index values ($R_{\mathrm{H}{\alpha}}$ less than about $1.05 \times 10^{-5}$) occupy the largest area of the parameter spaces.
The distributions of the middle-level $R_{\mathrm{H}{\alpha}}$ index values ($R_{\mathrm{H}{\alpha}}$ between about $1.05 \times 10^{-5}$ and $1.55 \times 10^{-5}$) in the parameter spaces are distinctly different for the main-sequence and giant samples.
For the main-sequence sample, the middle-level $R_{\mathrm{H}{\alpha}}$ values are mainly distributed in the usual main-sequence strip area in the $T_\mathrm{eff}$ -- $\log\,g$ parameter space and in a $\Lambda$-shaped strip area in the $T_\mathrm{eff}$ -- $\mathrm{[Fe/H]}$ parameter space;
while for the giant sample, the middle-level $R_{\mathrm{H}{\alpha}}$ values spread over an irregularly shaped area in the parameter spaces.
The distribution area of the upper-level $R_{\mathrm{H}{\alpha}}$ index values ($R_{\mathrm{H}{\alpha}}$ greater than about $1.55 \times 10^{-5}$) basically coincides with that of the middle-level $R_{\mathrm{H}{\alpha}}$ values in the $T_\mathrm{eff}$ -- $\mathrm{[Fe/H]}$ parameter space,
but do not exactly coincide in the $T_\mathrm{eff}$ -- $\log\,g$ parameter space, for both the main-sequence and giant samples;
this fact suggests that the upper-level and middle-level $R_{\mathrm{H}{\alpha}}$ values might have the same origin on stars, but are associated with stars at different evolution stages and ages. In general, the younger the star, the higher the activity level.

The minimum values of the $R_{\mathrm{H}{\alpha}}$ index are around $0.55 \times 10^{-5}$ for the main-sequence sample and less than about $0.60 \times 10^{-5}$ for the giant sample.
These minimum $R_{\mathrm{H}{\alpha}}$ index values are distributed in separate areas in the $T_\mathrm{eff}$ -- $\log\,g$ parameter space (one area for the main-sequence sample and two areas for the giant sample),
suggesting that the stellar atmospheric parameter $\log\,g$ has equal significance as the parameter $T_\mathrm{eff}$ in determining the bottom envelope of $R_{\mathrm{H}{\alpha}}$ distribution in the parameter space.

All the $I_{\mathrm{H}{\alpha}}$ and $R_{\mathrm{H}{\alpha}}$ index data obtained in this work are available in the online dataset of this paper. The web link and the data format of the dataset can be found in Section \ref{sec:dataset}.

The H$\alpha$ chromospheric activity distribution of F-, G-, and K-type stars obtained in this work from the LAMOST MRS data can be cross-compared with the results of the stellar chromospheric activity obtained from the LAMOST LRS data
(e.g., \citealt{2016A&A...594A..39F, 2020ApJS..247....9Z, 2021RAA....21..249B, 2022ApJS..263...12Z}),
the stellar photospheric activity and flare activity obtained from the light curve data of the \emph{Kepler} and \emph{TESS} missions
(e.g., \citealt{2012Natur.485..478M, 2015ApJS..221...18H, 2015EP&S...67...59M, 2017ApJ...834..207M, 2018ApJS..236....7H, 2019ApJS..244...37G, 2020AJ....159...60G, 2021ApJ...906...40G, 2021ApJ...906...72O, 2021ApJS..253...35T, 2021MNRAS.505L..79Y}),
and the stellar coronal activity obtained from the space-based X-ray observations (e.g., \citealt{2010ApJS..189...37E, 2011ApJ...743...48W, 2020ApJ...902..114W, 2020A&A...641A.136W, 2022A&A...664A.105F, 2022A&A...661A...6S}).
The algorithm and approach developed in this work for the H$\alpha$ activity indices can be applied to the single-exposure spectra and time-domain data of MRS (e.g., \citealt{2020RAA....20..167F, 2021RAA....21..292W, 2022A&A...663A.140L, 2022ApJ...928..180W, 2023ApJS..264...12H}),
from which the time evolution information of H$\alpha$ chromosphere activity can be revealed.
The dataset provided by this work can also be useful for assessing the impact of stellar activity to the radial velocity signals when searching for exoplanets, as well as when estimating space and atmospheric environment of exoplanets
(e.g. \citealt{2003A&A...403.1077K, 2007A&A...474..293B, 2009A&A...495..959B, 2011A&A...534A..30G, 2013ApJ...764....3R, 2016ApJ...832..112R}).

\backmatter

\bmhead{Acknowledgements}
Guoshoujing Telescope (the Large Sky Area Multi-Object Fiber Spectroscopic Telescope, LAMOST) is a National Major Scientific Project built by the Chinese Academy of Sciences. Funding for the project has been provided by the National Development and Reform Commission. LAMOST is operated and managed by the National Astronomical Observatories, Chinese Academy of Sciences.
This work made use of Astropy \citep{2013A&A...558A..33A, 2018AJ....156..123A}, PyAstronomy \citep{2019ascl.soft06010C}, PyDL, and SciPy \citep{2020NatMe..17..261V}.

\bmhead{Author contributions}
The study was carried out in collaboration of all authors. Han He performed the data analysis and wrote the manuscript with input from all coauthors.

\bmhead{Funding}
This research is supported by the National Key R\&D Program of China (2019YFA0405000) and the National Natural Science Foundation of China (11973059 and 12073001).
H.H. acknowledges the CAS Strategic Pioneer Program on Space Science (XDA15052200) and the B-type Strategic Priority Program of the Chinese Academy of Sciences (XDB41000000).
W.Z. and J.Z. acknowledge the support of the Anhui Project (Z010118169).

\bmhead{Data Availability}
The dataset generated during the current study is available in the online dataset of this paper (see Section \ref{sec:dataset} for the web link).

\section*{Declarations}

\bmhead{Informed Consent}
Informed consent was obtained from all individual participants included in the study.

\bmhead{Competing interests}
The authors declare no competing interests.

\begin{appendices}

\section{Distribution of the $\mathrm{[Fe/H]}$ values in the $T_\mathrm{eff}$ -- $\log\,g$ parameter space} \label{sec:feh_in_teff_logg_space}

\begin{figure}
  \centering
  \includegraphics[width=1.09\textwidth]{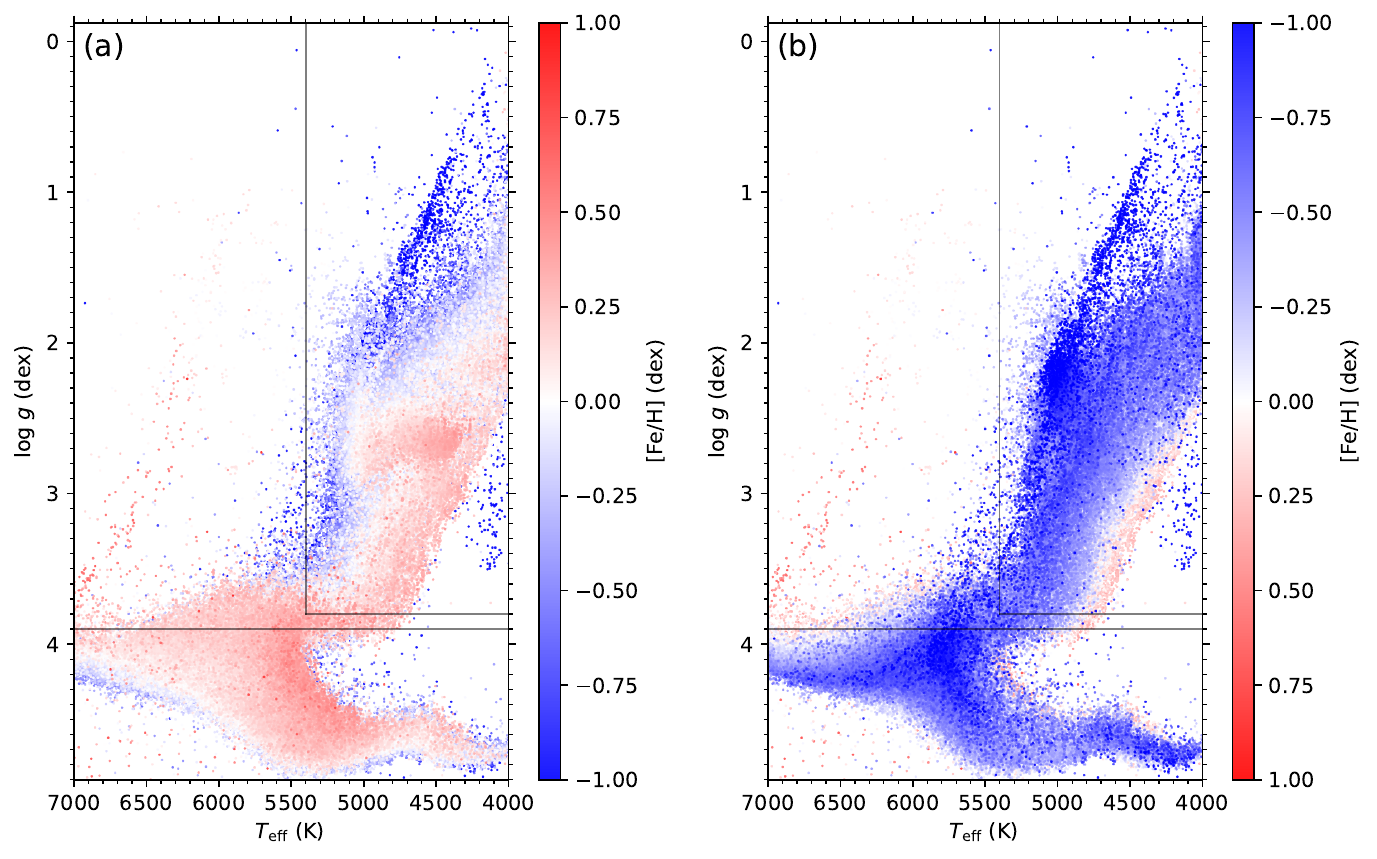}
  \caption{
    Distribution of the $\mathrm{[Fe/H]}$ values in the $T_\mathrm{eff}$ -- $\log\,g$ parameter space for the MRS spectra of F-, G-, and K-type stars analyzed in this work.
    The value of $\mathrm{[Fe/H]}$ is indicated by color scale (red for positive and blue for negative).
    In panel a, the data points are stacked in ascending order of their $\mathrm{[Fe/H]}$ values, with the maximum value at the highest layer and the minimum value at the lowest layer.
    In panel b, the data points are stacked in reverse order of their $\mathrm{[Fe/H]}$ values, with the minimum value at the highest layer and the maximum value at the lowest layer.
    The horizontal and vertical lines in the two panels are the dividing lines for the main-sequence and giant samples (same as in Figure \ref{fig:number_density_teff_vs_logg}).
  \label{fig:feh_in_teff_logg_space}}
\end{figure}

In Figure \ref{fig:feh_in_teff_logg_space}, we show the distribution of the $\mathrm{[Fe/H]}$ values in the $T_\mathrm{eff}$ -- $\log\,g$ parameter space for the MRS spectra of F-, G-, and K-type stars analyzed in this work. 
The values of the stellar atmospheric parameters ($T_\mathrm{eff}$, $\log\,g$, and $\mathrm{[Fe/H]}$) are provided by the LAMOST Stellar Parameter Pipeline (LASP) and are determined by matching the observed MRS spectra with the reference spectra. 
Panel a of Figure \ref{fig:feh_in_teff_logg_space} mainly shows the distribution of the positive $\mathrm{[Fe/H]}$ values (displayed in red) in the  $T_\mathrm{eff}$ -- $\log\,g$ parameter space, 
and panel b mainly shows the distribution of the negative $\mathrm{[Fe/H]}$ values (displayed in blue).  
The result exhibited in Figure \ref{fig:feh_in_teff_logg_space} does not suggest a degeneracy (apparent correlation) between the values of $\mathrm{[Fe/H]}$ and the values of $T_\mathrm{eff}$ (or $\log\,g$) provided by the LASP, demonstrating the reliability of the stellar parameter values.

\section{Relationship between the magnitudes of $\lowercase{v}{\sin}\lowercase{i}$ and $\Delta I_{\mathrm{H}{\alpha}}$} \label{sec:diff_I_index_vs_vsini}

We estimate the relationship between the magnitudes
of $v{\sin}i$ (projected rotational velocity) and $\Delta I_{\mathrm{H}{\alpha}}$ (increment of $I_{\mathrm{H}{\alpha}}$ caused by rotational broadening) by using the nine example spectra shown in Figure \ref{fig:example_mrs_halpha_spectra}.
These spectra are from different stellar types and at different H$\alpha$ activity levels and hence are representative.

For an input spectrum of H$\alpha$ line and a given $v{\sin}i$ value, the spectral profile after rotational broadening can be calculated by using the formula given by \citet{2008oasp.book.....G}.
For each of the spectra displayed in Figure \ref{fig:example_mrs_halpha_spectra}, we calculate a series of spectra after applying rotational broadening for a series of $v{\sin}i$ values from 0\,km/s to 30\,km/s,\footnote{The limb-darkening coefficient $\varepsilon$ is set to 0.6 in the calculation.}
and then evaluate $I_{\mathrm{H}{\alpha}}$ values for each of the calculated spectra.
The $\Delta I_{\mathrm{H}{\alpha}}$ values corresponding to the series of $v{\sin}i$ values are obtained by subtracting the $I_{\mathrm{H}{\alpha}}$ value of the original spectrum from the newly evaluated $I_{\mathrm{H}{\alpha}}$ values.
(Note that the value of $\Delta I_{\mathrm{H}{\alpha}}$ is positive for the H$\alpha$ absorption line profile and negative for the H$\alpha$ emission line profile.)

\begin{figure}
  \centering
  \includegraphics[width=1.018\textwidth]{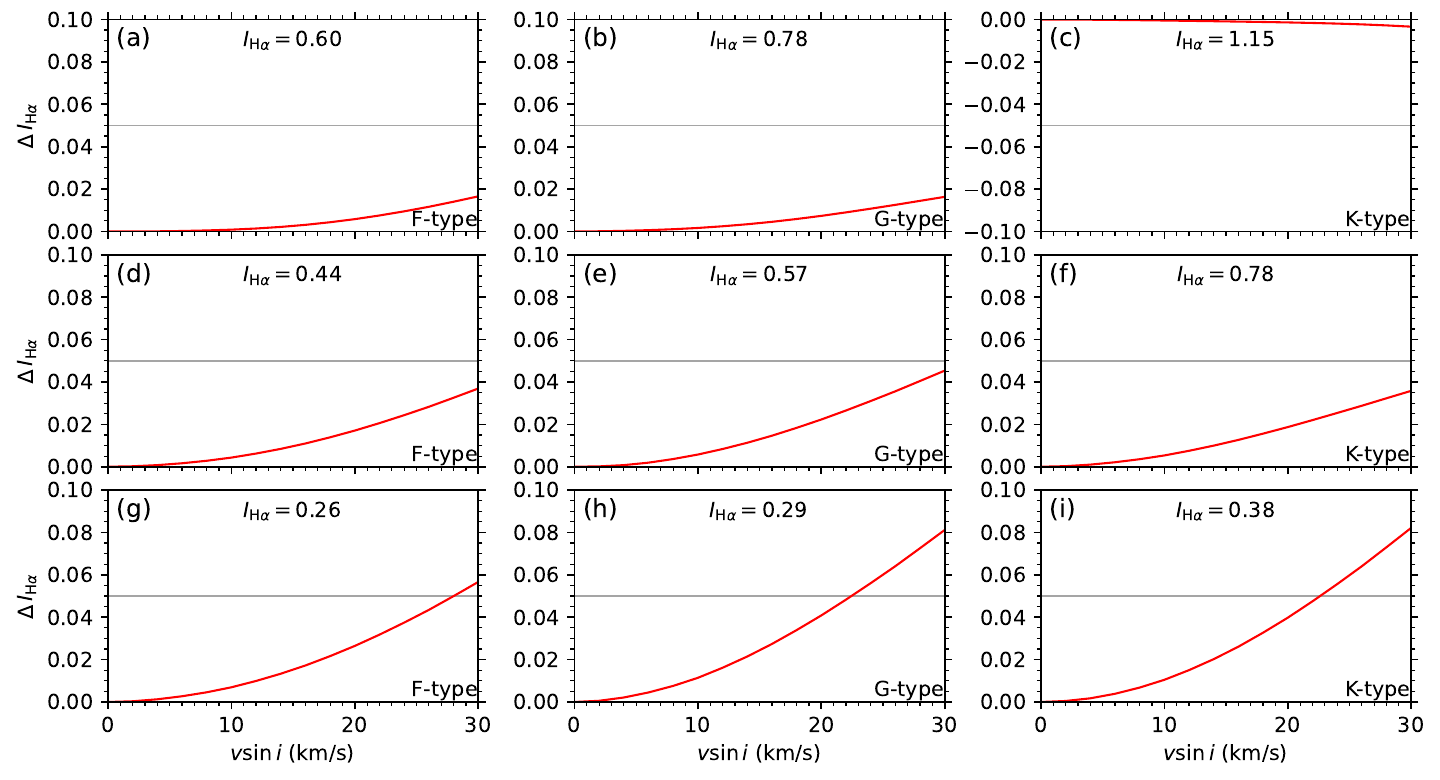}
  \caption{Plots of $v{\sin}i$ -- $\Delta I_{\mathrm{H}{\alpha}}$ relation (red curve) for the nine example spectra shown in Figure \ref{fig:example_mrs_halpha_spectra}.
  The horizontal line in the plots represents $|\Delta I_{\mathrm{H}{\alpha}}|=0.05$.
  The $I_{\mathrm{H}{\alpha}}$ values of the original spectra (corresponding to $v{\sin}i=0$\,km/s in the horizontal axis) are given in the plots.
  In panel c, $\Delta I_{\mathrm{H}{\alpha}}$ takes a negative value owing to the emission profile of the H$\alpha$ line.}
  \label{fig:I_index_increment_vs_vsini}
\end{figure}

The plots of $v{\sin}i$ -- $\Delta I_{\mathrm{H}{\alpha}}$ relation for all the nine example spectra are shown in Figure \ref{fig:I_index_increment_vs_vsini}.
It can be seen from Figure \ref{fig:I_index_increment_vs_vsini} that, for $v{\sin}i$ values less than 30\,km/s,
the absolute magnitude of $\Delta I_{\mathrm{H}{\alpha}}$ is generally below about $0.02 \thicksim 0.05$ for higher $I_{\mathrm{H}{\alpha}}$ values (panels a--f) and below about $0.06 \thicksim 0.08$ for lower $I_{\mathrm{H}{\alpha}}$ values (panels g--i).

It should be noted that a comparison between the $v{\sin}i$ values obtained by the LAMOST MRS and by the Apache Point Observatory Galactic Evolution Experiment (APOGEE; \citealt{2017AJ....154...94M}) shows that the $v{\sin}i$ by the MRS is slightly larger than that by the APOGEE (see the data release documents of LAMOST for more details),
and the original MRS spectra used for analyzing the $v{\sin}i$ -- $\Delta I_{\mathrm{H}{\alpha}}$ relation might have a small but non-zero intrinsic $v{\sin}i$.
Therefore, the $v{\sin}i$ -- $\Delta I_{\mathrm{H}{\alpha}}$ relation and the upper limit of $\Delta I_{\mathrm{H}{\alpha}}$ displayed in Figure \ref{fig:I_index_increment_vs_vsini} can be regarded as an estimate of magnitude rather than a precise result.

\section{Algorithm for $\delta\chi$ estimation} \label{sec:delta_chi_estimation}

The $\chi$ factor defined in Equation (\ref{equ:chi_definition}) is a function of stellar atmospheric parameters $T_\mathrm{eff}$, $\log\,g$, and $\mathrm{[Fe/H]}$,
thus the uncertainty of $\chi$ (i.e., $\delta\chi$) depends on the uncertainties of the three stellar atmospheric parameters $\delta T_\mathrm{eff}$, $\delta\log\,g$, and $\delta\mathrm{[Fe/H]}$.

We use $\delta\chi(T_\mathrm{eff})$, $\delta\chi(\log\,g)$, and $\delta\chi(\mathrm{[Fe/H]})$ to denote the uncertainties of $\chi$ caused by the uncertainties of the three stellar atmospheric parameters.
To estimate $\delta\chi(T_\mathrm{eff})$, we first calculate three $\chi$ values corresponding to $T_\mathrm{eff}$, $T_\mathrm{eff}-\delta T_\mathrm{eff}$, and $T_\mathrm{eff}+\delta T_\mathrm{eff}$, which are denoted by $\chi(T_\mathrm{eff})$, $\chi(T_\mathrm{eff}-\delta T_\mathrm{eff})$, and $\chi(T_\mathrm{eff}+\delta T_\mathrm{eff})$ respectively.
Then, the value of $\delta\chi(T_\mathrm{eff})$ is calculated by the following equation:
\begin{equation} \label{equ:chi_err_teff}
  \delta\chi(T_\mathrm{eff}) =
  \frac{\big|\chi(T_\mathrm{eff}-\delta T_\mathrm{eff}) -
  \chi(T_\mathrm{eff})\big| + 
  \big|\chi(T_\mathrm{eff}+\delta T_\mathrm{eff}) -
  \chi(T_\mathrm{eff})\big|}{2}.
\end{equation}
$\delta\chi(\log\,g)$ and $\delta\chi(\mathrm{[Fe/H]})$ can be estimated similarly.

The value of $\delta\chi$ is calculated from the values of $\delta\chi(T_\mathrm{eff})$, $\delta\chi(\log\,g)$, and $\delta\chi(\mathrm{[Fe/H]})$ by the following formula:
\begin{equation} \label{equ:chi_err}
  \delta\chi = \sqrt{
  \big[{\delta\chi}(T_\mathrm{eff})\big]^2 +
  \big[{\delta\chi}(\log\,g)\big]^2 +
  \big[{\delta\chi}(\mathrm{[Fe/H]})\big]^2
  }.
\end{equation}

\begin{figure}
  \centering
  \includegraphics[width=1.00\textwidth]{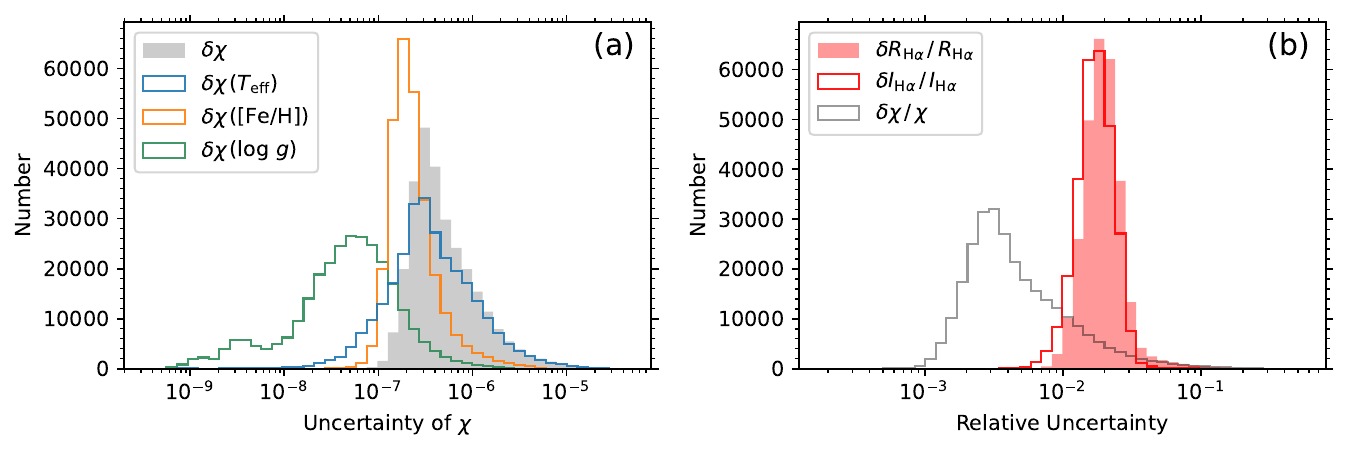}
  \caption{
  (a) Histograms of $\delta\chi$, $\delta\chi(T_\mathrm{eff})$, $\delta\chi(\mathrm{[Fe/H]})$, and $\delta\chi(\log\,g)$ for the MRS spectra analyzed in this work.
  (b) Histograms of $\delta R_{\mathrm{H}{\alpha}}/R_{\mathrm{H}{\alpha}}$, $\delta I_{\mathrm{H}{\alpha}}/I_{\mathrm{H}{\alpha}}$, and $\delta\chi/\chi$ for the MRS spectra analyzed in this work.
  }
  \label{fig:histogram_delta_chi}
\end{figure}

In Figure \ref{fig:histogram_delta_chi}a, we compare the histograms of $\delta\chi(T_\mathrm{eff})$, $\delta\chi(\log\,g)$, $\delta\chi(\mathrm{[Fe/H]})$, and $\delta\chi$ for the MRS spectra analyzed in this work. 
It can be seen from Figure \ref{fig:histogram_delta_chi}a that the value of $\delta\chi$ is mainly affected by $\delta\chi(T_\mathrm{eff})$,
and $\delta\chi(\log\,g)$ has the least effect on the $\delta\chi$ result.

In Figure \ref{fig:histogram_delta_chi}b, we compare the histograms of the relative uncertainties of $\chi$, $I_{\mathrm{H}{\alpha}}$, and $R_{\mathrm{H}{\alpha}}$ (i.e., $\delta\chi/\chi$, $\delta I_{\mathrm{H}{\alpha}}/I_{\mathrm{H}{\alpha}}$, and $\delta R_{\mathrm{H}{\alpha}}/R_{\mathrm{H}{\alpha}}$) for the MRS spectra analyzed in this work.
It can be seen from Figure \ref{fig:histogram_delta_chi}b that the relative uncertainty of $R_{\mathrm{H}{\alpha}}$ ($\delta R_{\mathrm{H}{\alpha}}/R_{\mathrm{H}{\alpha}}$) is mainly affected by $\delta I_{\mathrm{H}{\alpha}}/I_{\mathrm{H}{\alpha}}$, and for most MRS spectra $\delta\chi/\chi$ is much smaller than $\delta I_{\mathrm{H}{\alpha}}/I_{\mathrm{H}{\alpha}}$.
The influence of $\delta\chi/\chi$ on $\delta R_{\mathrm{H}{\alpha}}/R_{\mathrm{H}{\alpha}}$ is only prominent for a small number of spectra with relatively large uncertainties of $T_\mathrm{eff}$.

\end{appendices}

\end{document}